\documentclass[useAMS,usenatbib]{mn2e}

\usepackage{amssymb}
\usepackage{graphicx}
\usepackage{epsfig}

\newcommand{\ii}{\mathrm{i}}                                   
\newcommand{\dd}{\mathrm{d}}                                   

\begin{document}

\title[Corotation resonance in relativistic discs]{
	Corotation resonance and overstable oscillations in black-hole accretion discs: general-relativistic calculations
}    
  
\author[J.~Hor\'{a}k \& D.~Lai]{
	Ji\v{r}\'{\i}~Hor\'{a}k,$^1$ and Dong Lai$^2$\\ 
	$^1$Astronomical Institute of the Academy of Sciences, Bo\v{c}n\'{\i}~II 1401/1a, 141-31 Praha~4, CZ\\ 
	$^2$Department of Astronomy, Cornell University, Ithaca, NY 14853, USA
}
  
\date{Received; Accepted}

\pagerange{\pageref{firstpage}--\pageref{lastpage}} \pubyear{2013}
\newtheorem{theorem}{Theorem}[section]
\label{firstpage}

\maketitle


\begin{abstract}
We study the dynamics of spiral waves and oscillation modes in relativistic rotating discs around black holes. Generalizing the Newtonian theory, we show that wave absorption can take place at the corotation resonance, where the pattern frequency of the wave matches the background disc rotation rate. We derive the general relativistic expression for the disc vortensity (vorticity divided by surface density), which governs the behaviour of density perturbation near corotation. Depending on the gradient of the generalized disc vortensity, corotational wave absorption can lead to the amplification or damping of the spiral wave. We apply our general theory of relativistic wave dynamics to calculate the non-axisymmetric inertial-acoustic modes (also called p-modes) trapped in the inner-most region of a black hole accretion disc. Because general relativity changes the profiles of the radial epicyclic frequency and disc vortensity near the inner disc edge close to the black hole, these p-modes can become overstable under appropriate conditions. We present the numerical results of the frequencies and growth rates of p-modes for various black hole spin and model disc parameters (the surface density profile and sound speed), and discuss their implications for understanding the enigmatic high-frequency quasi-periodic oscillations observed in black hole X-ray binaries.
\end{abstract}



\section{Introduction}
High-frequency quasi-periodic oscillations (HFQPOs) in the X-ray fluxes have been observed from a number of black hole (BH) X-ray binaries since the late 1990s. Their frequencies (40-450~Hz) are comparable to the orbital frequency of a test mass close to the BH (within 10 gravitational radii) and do not vary significantly in response to sizeable (factors of 5 or more) changes in the luminosity. Compared to the low-frequency QPOs, the HFQPOs are weaker and more transient, and are only observed in the `intermediate state' (or `steep power-law state') of the X-ray binaries \citep[for reviews, see][]{Remillard+McClintock2006, Belloni+2012}. In addition, several systems show pairs of QPOs with frequency ratios close to 3:2. It is generally recognized that HFQPOs may provide an important information about the BH (mass and spin) and about the dynamics of inner-most accretion flows \citep[e.g.,]{Torok+2005}.

The physical origin of HFQPOs is currently unclear. A number of ideas/models have been suggested or studied to various degrees of sophistication. For example, in the spot models \citep{Stella+1999} the HFQPOs arise from the Doppler modulation of the radiation from spots orbiting in the inner part of the BH accretion disc. But it is not clear how the position of the spot (a free parameter in these models) is determined and how the spot can survive the differential rotation of the disc. Another class of models identify HFQPOs with various oscillation modes of a finite, pressure-supported accretion torus \citep{Rezzolla+2003, Blaes+2006}. However, it is not clear that the accretion flow can be approximated by a torus and how the position and size of the torus are determined. The harmonic relation between the observed frequencies led \citet{Abramowicz+Kluzniak2001} to suggest that the HFQPOs are a result of a nonlinear resonance \citep[see also][]{Kluzniak+Abramowicz2002}. However, detailed calculations of the resonant coupling between the epicyclic modes in slender tori indicates that such resonance is very weak \citep{Horak2008}.

Another class of theoretical models is based on the relativistic diskoseismology \citep[for reviews, see][]{Kato2001, Kato2008, Wagoner2008}. \citet{Okazaki+1987} first noticed that general-relativistic effects allow oscillation modes to be trapped in the inner region of a BH accretion disc. The g-modes (also called inertial or inertial-gravity modes) have attracted most attention, because their existence does not require a reflective inner disc boundary. These oscillations have at least one node in the vertical direction and their restoring force results from rotation and gravity. Unfortunately, the non-axisymmetric g-modes are either damped due to corotation resonance \citep{Kato2003, Li+2003, Tsang+Lai2009a} or have frequencies too high compared to HFQPOs \citep{Silbergleit+Wagoner2008}. The axisymmetric g-mode trapped around the maximum of the radial epicyclic frequency may account for some single HFQPO frequencies \citep[see][]{Wagoner2012} and several studies suggested that they may be resonantly excited in the warped or eccentric discs \citep{Kato2008, Ferreira+Ogilvie2008, Henisey+2009, Kato2012}. However, the self-trapping property of $g$-modes can easily destroyed by a weak (sub-thermal) disc magnetic field \citep{Fu+Lai2009} and turbulence \citep{Arras+2006, Reynolds+Miller2009}.

Perhaps more promising are the disc p-modes (also called inertial acoustic modes). These modes represent nearly horizontal oscillations with almost no vertical structure, whose main restoring force results from pressure gradients. They are trapped between the inner boundary of the disc and the inner Lindblad resonance (ILR), where the condition $\omega-m\Omega=-\kappa$ is satisfied (here, $\omega$ is the mode frequency, $\Omega$ is the disc rotation rate, $m$ is the azimuthal mode number and $\kappa$ is the radial epicyclic frequency).  Because of their simple two-dimensional (2D) structure, the basic wave properties of the p-modes (e.g., propagation diagram) are not strongly affected by disc magnetic fields \citep{Fu+Lai2009} and are likely robust in the presence of disc turbulence \citep[see][]{Arras+2006, Reynolds+Miller2009}. Recently, \citet{Lai+Tsang2009}, \citep[see also][]{Tsang+Lai2008, Tsang+Lai2009c} showed that the non-axisymmetric p-modes can naturally grow due to the \textit{corotational instability}.  This instability arises because of two effects: (1) since the waves inside the ILR carry negative energies while those outside the outer Lindblad resonance (OLR) (where $\omega-m\Omega=\kappa$) carry positive energies; the leakage of the p-waves through the corotation barrier (between ILR and OLR) leads to mode growth. (2) More importantly, when the vortensity of the disc flow \footnote{This applies to barotropic discs, for which the pressure is a unique function of the density. See \citet{Tsang+Lai2009c} for non-barotropic discs.}, $\zeta=\kappa^2/(2\Omega\Sigma)$ (where $\Sigma$ is the surface density), has a positive slope at the corotation radius (where $\omega=m\Omega$), wave absorption at the CR leads to amplification of the trapped p-mode. The non-linear evolution of overstable p-modes was recently studied by \citet{Fu+Lai2013} using 2D simulations.

The strong gravity is a key factor for the corotational instability. In Newtonian theory, the vortensity condition $(d\zeta/dr)_\mathrm{CR} > 0$ is not satisfied for uniform or smooth surface density profile, and the p-modes are strongly damped. In general relativity (GR), the radial epicyclic frequency $\kappa$ reaches a maximum before decreasing to zero at the innermost stable circular orbit (ISCO). This causes a non-monotonic behaviour in the vortensity profile, making mode growth possible. The previous calculations by \citet{Lai+Tsang2009} and \citet{Tsang+Lai2009c} adopted the pseudo-Newtonian potential of \citet{Paczynsky+Viita1980} to mimic the effects of strong gravity. But a quantitative description of the corotational instability and overstable $p$-modes of BH accretion discs requires full GR. Note that although oscillation modes of relativistic discs were studied in several previous works \citep[see][for reviews]{Ortega-Rodriguez+2008, Wagoner2008}, the important role of CR was overlooked in these works.

In this paper, we formulate the theory of the corotational instability in the framework of the GR. In Section~\ref{sec:theory}, we derive the governing equations for vertically integrated perturbations of accretion discs in a general axisymmetric spacetime. These equations are then solved numerically for the particular case of the Kerr spacetime and results are presented in Section~\ref{sec:kerr}. Section~\ref{sec:discussion} is devoted to the discussion of our results and their applications to models of HFQPOs.


\section{Theory}
\label{sec:theory}


\subsection{Preliminaries}

We consider a fluid disc surrounding a compact object of mass $M$ generating a stationary, axisymmetric space-time. The nonzero components of the corresponding metric tensor are $g_{tt}$, $g_{t\phi}$, $g_{rr}$, $g_{\theta\theta}$ and $g_{\phi\phi}$. We employ the $(-,+,+,+)$-signature of the metric and use the units where $G=c=M=1$ throughout the paper.

The components of the inverse metric tensor are given by $g^{tt} = -g_{\phi\phi}/\mathcal{R}^2$, $g^{t\phi} = g_{t\phi}/\mathcal{R}^2$, $g^{\phi\phi} = -g_{tt}/\mathcal{R}^2$, $g^{rr} = 1/g_{rr}$ and $g^{\theta\theta} = 1/g_{\theta\theta}$, in which $\mathcal{R}^2 = g_{t\phi}^2 - g_{tt}g_{\phi\phi}$ (note that $\mathcal{R}\rightarrow r\sin\theta$ as $r\rightarrow\infty$). We assume that the fluid in the unperturbed disc is in purely orbital motion, neglecting radial infall -- this is valid away from the disc inner edge. The contravariant and covariant components of the four-velocity of the fluid are given by
\begin{equation}
	u^\alpha = u^t(\delta^\alpha_t + \Omega\delta^\alpha_\phi), 
	\quad
	u_\beta = u_t(\delta^t_\beta - \ell\delta^\phi_\beta),
\end{equation}
where $\Omega = u^\phi/u^t$ is the angular velocity measured by a distant observer and $\ell = -u_\phi/u_t$ measures the specific angular momentum. They are mutually related by
\begin{equation}
	\Omega = \frac{g^{t\phi} - \ell g^{\phi\phi}}{g^{tt} - \ell g^{t\phi}}, 
	\quad
	\ell = -\frac{g_{t\phi} + \Omega g_{\phi\phi}}{g_{tt} + \Omega g_{t\phi}}
\end{equation}
Finally, it follows from the normalization of the four-velocity, $u^\alpha u_\alpha = -1$, that
\begin{equation}
	u^t u_t (1-\ell\Omega) = -1.
\end{equation}


\subsection{Perturbation equations}

Dynamics of the disc follows from the Euler equation (we ignore effects of viscosity),
\begin{equation}
	\nabla_\alpha T^\alpha_\beta = 0, 
	\quad
	T^\alpha_\beta = (e+p) u^\alpha u_\beta + p\delta^\alpha_\beta,
\end{equation}
and from the mass continuity equation,
\begin{equation}
	\nabla_\alpha\left(\rho u^\alpha\right) = 
	\frac{1}{\sqrt{-g}} \partial_\alpha\left(\sqrt{-g}\rho u^\alpha\right) = 0.
\end{equation}
In the above equations, $T^\alpha_\beta$ is the stress-energy tensor, $e$, $\rho$, $p$ are the energy density, rest-mass density and pressure, respectively, and $g = -g_{rr} g_{\theta\theta}\mathcal{R}^2$ is the determinant of the metric tensor.

Perturbing the Euler equation we obtain
\begin{eqnarray}
	\left\{u^\alpha\nabla_\alpha(\delta e + \delta p) + 
	\nabla_\alpha\left[(e+p)\delta u^\alpha\right]\right\} u_\beta +
	&\phantom{=}& \nonumber \\ 
	(\delta e + \delta p) a_\beta +
	(e+p)\delta a_\beta + \nabla_\beta \delta p &=& 0,
	\label{eq:Euler-pert-1}
\end{eqnarray}
where $a_\beta = u^\alpha\nabla_\alpha u_\beta$ is four-acceleration of the flow. The notation $\delta$ stands for Eulerian perturbation. The perturbation of $a_\beta$ is given by
\begin{equation}
	\delta a_\beta = u^\alpha \delta u_{\beta,\alpha} + u_{\beta,\alpha}\delta u^\alpha - 
	g_{\mu\nu,\beta}u^\mu \delta u^\nu.
\end{equation}
Equation (\ref{eq:Euler-pert-1}) describes the evolution of all components of the four-velocity perturbation. In fact, they are not independent
because the four-velocity of the perturbed flow always has to satisfy normalization condition and therefore
\begin{equation}
	\delta u_t = -\Omega \delta u_\phi, \quad
	\delta u^t = \ell \delta u^\phi.
\end{equation}

Since the unperturbed flow is stationary and axisymmetric, we further assume that perturbations of all quantities depend on time and
the azimuth as $\delta \propto \exp[-\ii(\omega t -  m\phi)]$. Contracting equation (\ref{eq:Euler-pert-1}) with the vector $w^\beta = \delta^\beta_\phi + \ell \delta^\beta_t$, we find
\begin{equation}
	(e+p)\left[\ii\tilde{\omega} \frac{\delta u_\phi}{u_t} - u_t\ell_{,k}\delta u^k\right] +
	\ii\tilde{m}\delta p = 0,
	\label{eq:Euler-azimuth}
\end{equation}
where 
\begin{equation}
	\tilde{\omega} = \omega - m\Omega,\quad
	\tilde{m}=m-\ell\omega.
\end{equation}
The remaining poloidal components (in what follows denoted by Latin indices $i=r,\theta$) are
\begin{equation}
	(\delta e + \delta p) a_i - (e+p)\left[\ii\tilde{\omega} u^t\delta u_i +
	A_i\frac{\delta u_\phi}{u_t}\right] + \delta p_{,i} = 0,
	\label{eq:Euler-poloidal}
\end{equation}
where
\begin{eqnarray}
	A_i &=& -u_t^2\left[(g^{tt}-\ell g^{t\phi})\Omega_{,i} + 
	(g^{\phi\phi}-\Omega g^{t\phi})\ell_{,i}\right]
	\nonumber \\
	&=& \frac{\Omega_{,i}}{1-\ell\Omega} + 
	\frac{u_t^3}{u^t}\frac{\ell_{,i}}{\mathcal{R}^2}.
	\label{eq:Ai}
\end{eqnarray}
Finally, perturbing the continuity equation we find that
\begin{equation}
	\frac{1}{\sqrt{-g}}\partial_k\left(\sqrt{-g}\rho\,\delta u^k\right)
	-\ii\rho\frac{\tilde{m}}{\mathcal{R}^2}\frac{u_t^2}{u^t}\frac{\delta u_\phi}{u_t}
	-\ii\tilde{\omega}u^t\delta\rho = 0.
	\label{eq:continuity}
\end{equation}


\subsection{Perturbation equations for vertically integrated discs}

We consider a geometrically thin disc made of barotropic fluid. The half-thickness of the disc in the $\theta$-coordinate is $\theta_\mathrm{m}\ll 1$, i.e.\ the disc region correspond to the interval $\pi/2-\theta_\mathrm{m}<\theta<\pi/2+\theta_\mathrm{m}$. The angular momentum profile is therefore approximately Keplerian, $\ell\approx \ell_\mathrm{K}$, and the fluid follows approximately geodesics and thus $a_r\approx a_\theta\approx 0$. The local speed of sound is always negligible with respect to the speed of light, hence $e+p\approx\rho$. Both the four-velocity of the unperturbed flow and the metric tenor vary slowly through the disc thickness and can be approximated by their values in the equatorial plane,
\begin{equation}
	u^\alpha(r,\theta)\approx u^\alpha(r,\pi/2),
	\quad
	g_{\alpha\beta}(r,\theta)\approx g_{\alpha\beta}(r,\pi/2).
\end{equation}
in the region $\pi/2-\theta_\mathrm{m}<\theta<\pi/2+\theta_\mathrm{m}$. 

In this paper, we concentrate on the disc p-modes (also called `inertial-acoustic modes'). These modes have zero node in the $\theta$-direction in their wavefunctions, and generally have a weak dependence on $\theta$ across the disc thickness. Hence, we replace also the velocity perturbation $\delta u^i$ and the enthalpy perturbation $\delta h=\delta p/\rho$ with their equatorial values. The $\theta$-component of equation (\ref{eq:Euler-poloidal}) with $\delta h_{,\theta}\approx 0$ implies that $\delta u_\theta$ vanishes. Then we integrate equations (\ref{eq:Euler-azimuth}), (\ref{eq:Euler-poloidal}) and (\ref{eq:continuity}) over $\theta$ and obtain
\begin{eqnarray}
	-\ii\tilde{\omega}\frac{\delta u_\phi}{u_t} + u_t \ell_{,r}\delta u^r - \ii\tilde{m}\delta h &=& 0,
	\label{eq:azimuthal-2}
	\\
	-\ii\tilde{\omega} u^t\delta u_r - A\frac{\delta u_\phi}{u_t} + \delta h_{,r} &=& 0,
	\label{eq:radial-2}
	\\
	-\ii\tilde{\omega} u^t \frac{\Sigma}{\bar{c}_\mathrm{s}^2}\delta h + 
	\frac{1}{\sqrt{-g_3}}\partial_r\left(\sqrt{-g_3}\Sigma\,\delta u^r\right)
	&\phantom{=}& \nonumber \\
	-\ii\Sigma\tilde{m}\frac{1}{\mathcal{R}^2}\frac{u_t^2}{u^t}\frac{\delta u_\phi}{u_t} &=& 0,
	\label{eq:continuity-2}
\end{eqnarray}
where $g_3 = g/g_{\theta\theta} = -g_{rr}\mathcal{R}^2$, and
\begin{equation}
	\Sigma = \int_{-\theta_\mathrm{m}}^{\theta_\mathrm{m}}\rho\sqrt{g_{\theta\theta}}\dd\theta,
	\quad
	\bar{c}_\mathrm{s}^2 = \frac{\Sigma}{\int_{-\theta_\mathrm{m}}^{\theta_\mathrm{m}}
	(\rho/c_\mathrm{s}^2)\sqrt{g_{\theta\theta}}\dd\theta}
\end{equation}
are the surface density and vertically averaged sound speed squared, respectively, $A = A_r$ is evaluated at the equatorial plane.

Equations (\ref{eq:azimuthal-2})--(\ref{eq:continuity-2}) are relativistic version of equations (6)--(8) of \citet{Lai+Tsang2009}. Using the azimuthal equation (\ref{eq:azimuthal-2}) to eliminate the azimuthal velocity perturbation, we have
\begin{equation}
	\frac{\delta u_\phi}{u_t} = -\frac{1}{\tilde{\omega}}\left(
	\tilde{m}\delta h + \ii\ell_{,r}u_t\delta u^r\right).
\end{equation}
Substituting this into equations (\ref{eq:radial-2}) and (\ref{eq:continuity-2}), we obtain
\begin{eqnarray}
	\delta h_{,r} &=& -\frac{1}{\tilde{\omega}}\left[
	\tilde{m}A\delta h + \ii D u^t\delta u_r\right], 
	\label{eq:h}
	\\
	\delta u_{r,r} &=& \ii\tilde{\omega}g_{rr} u^t\left[
	\frac{1}{\bar{c}_\mathrm{s}^2}-\left(\frac{\tilde{m}}{\tilde{\omega}}
	\frac{u_t}{\mathcal{R}u^t}\right)^2\right]\delta h 
	\nonumber \\ &\phantom{=}&
	+ \left[\frac{\tilde{m}}{\tilde{\omega}}\frac{u_t^3}{\mathcal{R}^2u^t}\ell_{,r} -
	\partial_r\ln\left(\frac{\sqrt{-g_3}\Sigma}{g_{rr}}\right)\right]\delta u_r,
	\label{eq:u}
\end{eqnarray}
with 
\begin{equation}
	D=\kappa^2-\tilde{\omega}^2, 
\end{equation}
and $\kappa$ is the radial epicyclic frequency and is given by
\begin{equation}
	\kappa^2 = \frac{A}{g_{rr}}\frac{u_t}{u^t}\ell_{,r}.
	\label{eq:kappa-A}
\end{equation}
In Section 3, equations (\ref{eq:h}) and (\ref{eq:u}) will be integrated numerically with appropriate boundary conditions in seeking for global disc p-modes.


\subsection{Wave equation and resonances}

Isolating the velocity perturbation from equation (\ref{eq:h}), we obtain
\begin{equation}
	\delta u_r = \frac{\ii}{u^t D}\left[\tilde{\omega} \delta h_{,r} + \tilde{m}A\delta h\right].
	\label{eq:h2u}
\end{equation}
Substituting this into equation (\ref{eq:u}) and after some algebra, we arrive at the wave equation for the enthalpy perturbation:
\begin{eqnarray}
	\delta h_{,rr} &+& \partial_r\ln\left(\frac{\sqrt{-g_3}\Sigma}{g_{rr}D(u^t)^2}\right)\delta h_{,r}
	\nonumber \\
	&+&\Big\{\frac{\tilde{m}}{\tilde{\omega}}A\partial_r\ln
	\left(\frac{\sqrt{-g_3}}{g_{rr}}\frac{\tilde{m}A\Sigma}{D u^t}\right) 
	\nonumber \\
	&\phantom{+}& -g_{rr}(u^t)^2\left[\frac{D}{\bar{c}_\mathrm{s}^2} + \left(\frac{u_t}{u^t}\right)^2
	\frac{\tilde{m}^2}{\mathcal{R}^2}\right]\Big\}\delta h = 0.
	\label{eq:master}
\end{eqnarray}
The first-order term ($\propto \delta h_{,r}$) can be eliminated by introducing a new variable
\begin{equation}
	\eta = S^{-1/2} \delta h,
	\quad
	S = \frac{g_{rr}D(u^t)^2}{\sqrt{-g_3}\Sigma}.
\end{equation}
Then equation (\ref{eq:master}) becomes
\begin{equation}
	\left[\frac{\dd^2}{\dd r^2} - V_\mathrm{eff}(r)\right]\eta(r) = 0,
	\label{eq:schrodinger}
\end{equation}
where
\begin{eqnarray}
	V_\mathrm{eff} &=& 
	g_{rr}(u^t)^2\left[\frac{D}{\bar{c}_\mathrm{s}^2} + 
	\left(\frac{u_t}{u^t}\right)^2\frac{\tilde{m}^2}{\mathcal{R}^2}\right] 
	\nonumber \\ &\phantom{=}&
	- \frac{\tilde{m}A}{\tilde{\omega}}\partial_r\ln
	\left(\frac{\sqrt{-g_3}}{g_{rr}}\frac{\tilde{m}A\Sigma}{D u^t}\right) +
	S^{1/2}\partial_r^2 S^{-1/2}.
	\label{eq:master-2}
\end{eqnarray}

\begin{figure*}
	\includegraphics[width=0.48\textwidth]{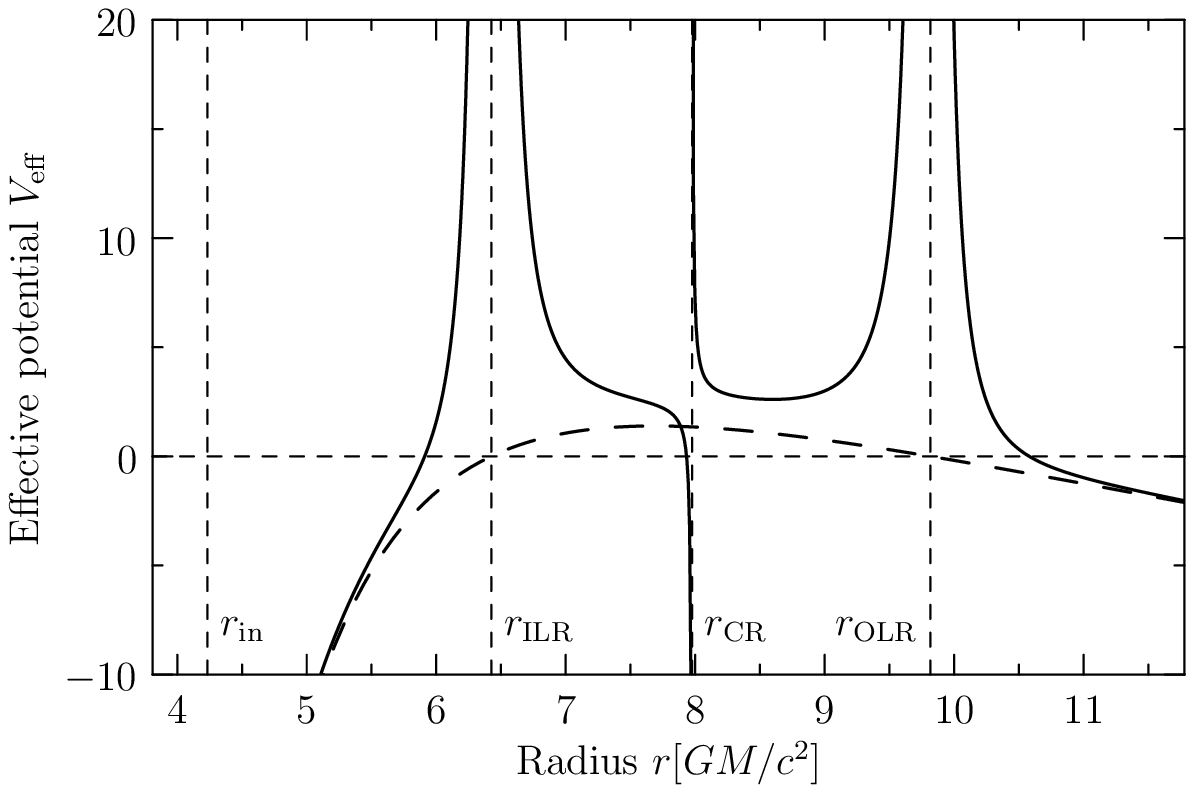}
	\hfill
	\includegraphics[width=0.48\textwidth]{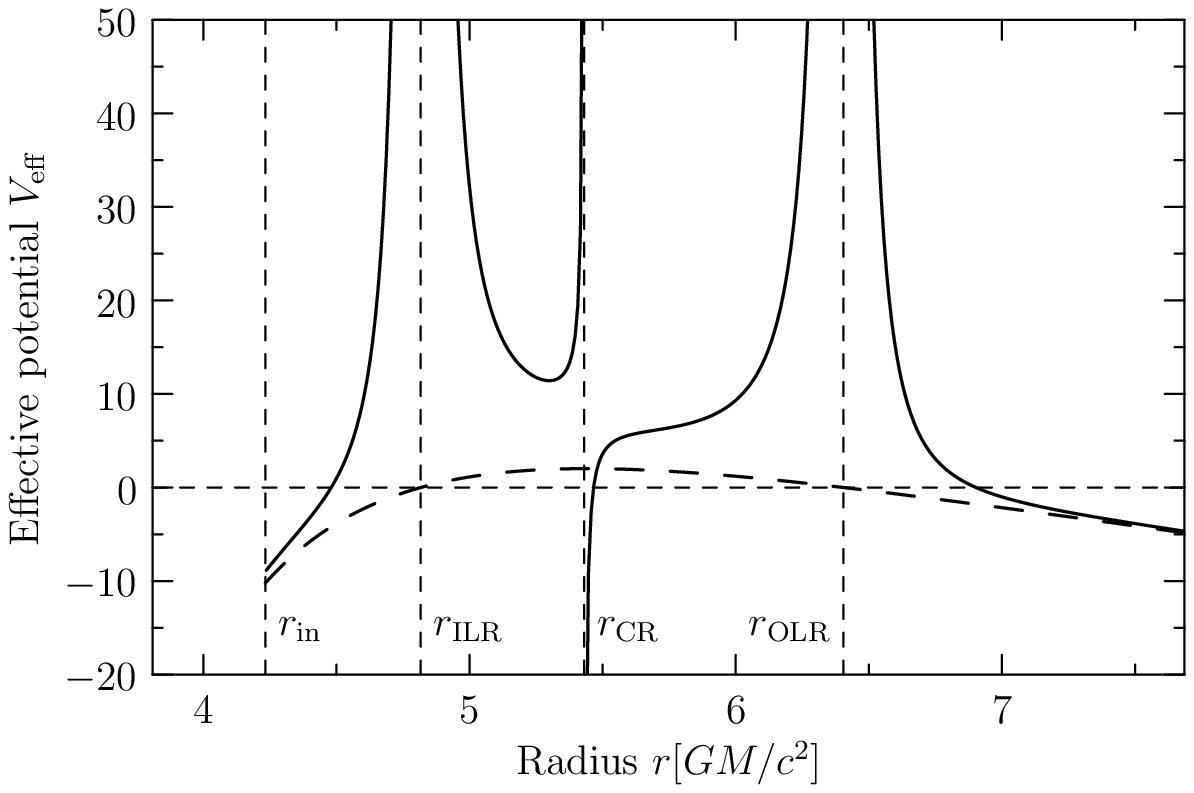}
	\caption{
		The effective potential $V_\mathrm{eff}$ for the wave propagation introduced in equations (\ref{eq:schrodinger}) and (\ref{eq:master-2}). The regions of wave propagation and evanescent regions correspond to $V_\mathrm{eff}<0$ and $V_\mathrm{eff}>0$, respectively. The dashed vertical lines denote four important radii: the inner edge of the disc $r_\mathrm{in}$, the corotation resonance $r_\mathrm{CR}$ and the inner and outer Lindblad resonances, $r_\mathrm{ILR}$ and $r_\mathrm{OLR}$. The solid line corresponds to the full expression (\ref{eq:master-2}) and the dashed line denotes the WKBJ approximation $V_\mathrm{eff}\approx -k^2$ valid far from the    resonances. The $p$-wave can freely propagate inside $r_{\rm ILR}$, outside $r_{\rm OLR}$ and in a small Rossby-wave zone adjected to  the corotation resonance. For a given set of disc and BH parameters, depending on the wave frequency $\omega$, the slope of the relativistic vortensity at $r_{\rm CR}$ can be either positive or negative, and the Rossby-wave zone is located either outside or inside $r_{\rm CR}$ In the former case, waves of positive energy are absorbed, leading to the overstability of the oscillations. The figure is made for $m=2$ waves in the disc with constant density and sound speed $\bar{c}_\mathrm{s} = 0.1 r\Omega_\mathrm{K}$, surrounding a BH with spin $a=0.5$. The left-hand and right-hand panels correspond to the wave-frequency of $0.4m\Omega_\mathrm{ISCO}$ and$0.7m\Omega_\mathrm{ISCO}$, respectively.
	}
	\label{fig:effpot}
\end{figure*}

Equation (\ref{eq:schrodinger}) resembles the stationary Schr\"odinger equation with effective potential $V_\mathrm{eff}$. The regions of wave propagation are given by the condition $V_\mathrm{eff}(r)<0$. For thin discs ($\bar{c}_\mathrm{s}\ll r\Omega$), this potential can be approximated by
\begin{equation}
	V_\mathrm{eff}\simeq  -k^2 + k\,\partial_r^2 \left(\frac{1}{k}\right) -
	\frac{\tilde{m}A}{\tilde{\omega}}\partial_r\ln
	\left(\frac{\sqrt{-g_3}}{g_{rr}}\frac{\tilde{m}A\Sigma}{D u^t}\right),
	\label{eq:master-3}
\end{equation}
where
\begin{equation}
	k = g_{rr}^{1/2} u^t \frac{\sqrt{-D}}{\bar{c}_\mathrm{s}},
	\quad
	\mathrm{Re}\,k \geq 0.
	\label{eq:k}
\end{equation}

An example of the effective potential is shown in Fig.~\ref{fig:effpot}. The singularities of $V_{\rm eff}$ occur at points where $D=0$ and $\tilde{\omega}=0$, corresponding to the Lindblad resonances and CR, respectively. As in the non-relativistic case, the ILR (where $\tilde\omega=-\kappa$) and the OLR (where $\tilde\omega=\kappa$) are only apparent singularities \citep[e.g.][see below]{Goldreich+Tremaine1979, Tsang+Lai2008}. The Lindblad resonances are turning points of the wave equation, separating the regions of wave propagation from those where waves become evanescent. The CR is a true singularity, where energy exchange between the wave and the background flow can take place.

The nonaxisymmetric p-modes (inertial-acoustic modes), which are the subject of this paper, are partially trapped  between the inner edge of the disc and the ILR. The CR at $r=r_\mathrm{CR}$ is surrounded by the two Lindblad resonances at $r=r_\mathrm{ILR}$ and $r_\mathrm{OLR}$. The p-mode can penetrate the corotation barrier and leak out as an outgoing wave in the wave zone $r>r_{\rm OLR}$. Before solving for these global p-modes in Section 3, we discuss below the behaviour of the fluid perturbations around the Lindblad resonances and CR.

\begin{figure*}
	\includegraphics[width=0.48\textwidth]{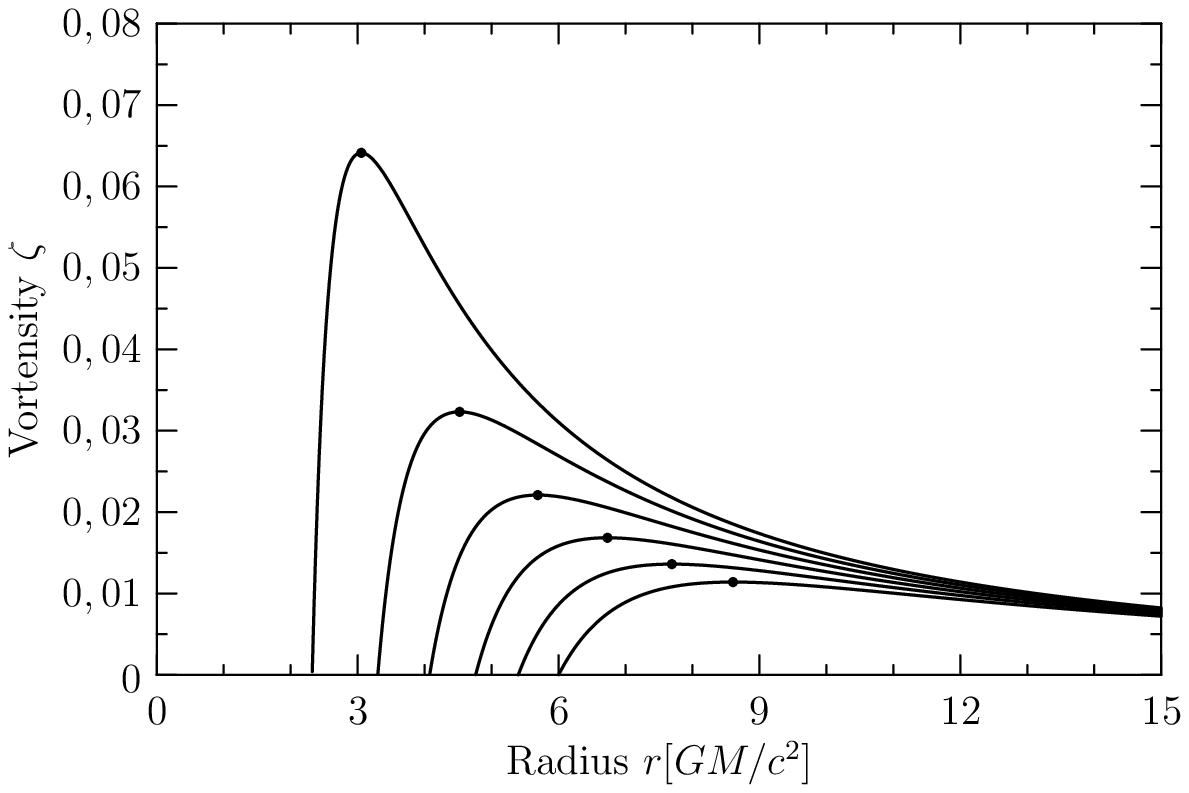}
	\hfill
	\includegraphics[width=0.48\textwidth]{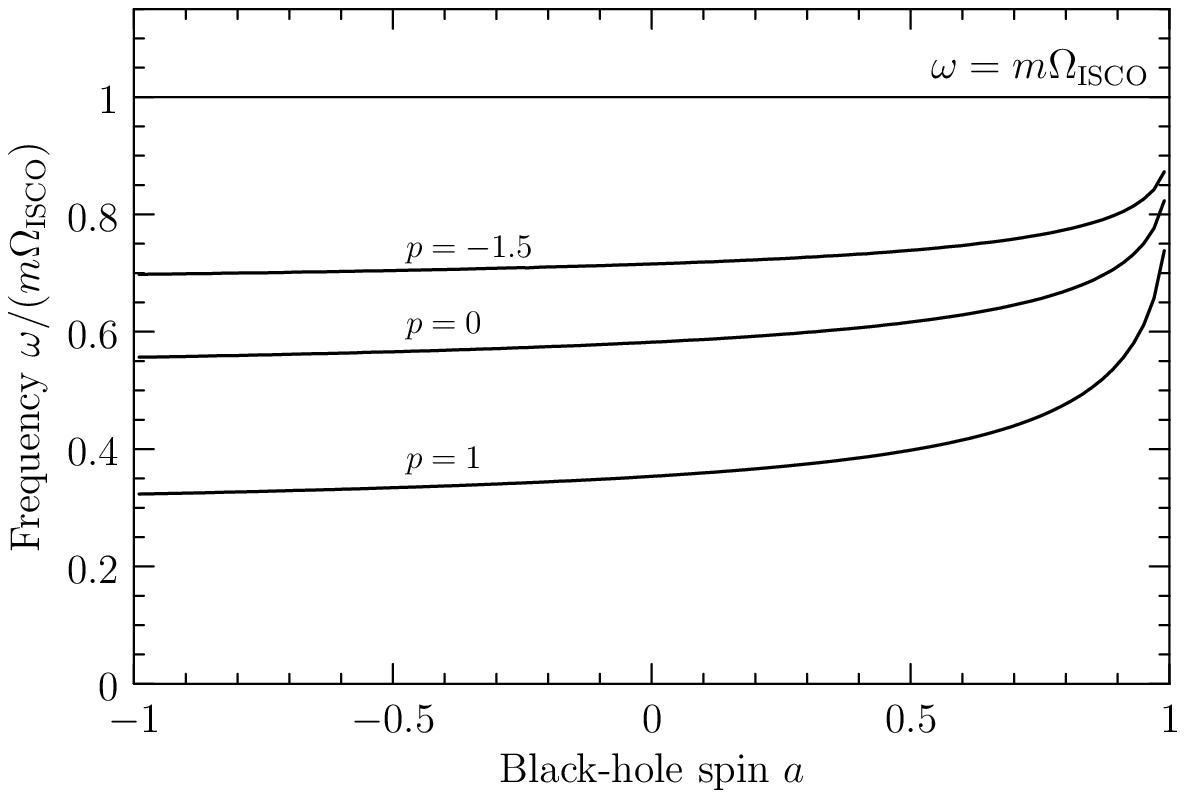}
	\caption{
		Left: vortensity profiles for a relativistic disc with constant surface density profile ($p=0$) surrounding a BH with $a = 0, 0.18, 0.36, 0.54, 0.72, 0.9$ (from the bottom curve to the top one). Right: the minimum frequency of unstable p-mode $\omega_\mathrm{min}$ due to wave absorption at the corotation resonance as a function of the BH spin for $p=-1.5,~0$ and $1.5$.
	}
	\label{fig:vortensity}
\end{figure*}

\subsection{WKBJ approximation}

Far from the Lindblad resonances and CRs, $(-k^2)$ is the dominant term on the right-hand side of equation (\ref{eq:master-3}). Hence, $k$ is the radial wavevector and the WKBJ approximation of the solution is
\begin{eqnarray}
	\delta h = S^{1/2}\eta &=& \left(\frac{S}{k}\right)^{1/2}\Big[
	A_\mathrm{-}\exp\left(-\ii\int^r k \dd r\right) 
	\nonumber \\ 
	&+& A_\mathrm{+}\exp\left(\ii\int^r k \dd r\right)\Big]
\end{eqnarray}
with $A_{+}$ and $A_{-}$ being constants. For $r>r_\mathrm{OLR}$ they correspond to the amplitudes of the ingoing and outgoing (with respect to the central BH; we assume that $Re\omega \geq 0$) and vice versa for $r<r_\mathrm{ILR}$ (note that for $r_{\rm ILR}$, the group velocity of the wave is opposite to the phase velocity). The wavevector $k$ and frequency $\omega$ are connected by the WKBJ dispersion relation
\begin{equation}
	\tilde{\omega}^2 = \kappa^2 + \frac{\bar{c}_\mathrm{s}^2 k^2}{g_{rr}(u^t)^2}.
\end{equation}

\subsection{Regularity of the solution at Lindblad resonances}

Close to the Lindblad resonance at $r=r_\mathrm{LR}$ we approximate $D$ as $D \approx D^\prime_\mathrm{LR}(r-r_\mathrm{LR})$ ($r_\mathrm{LR}$ is either $r_\mathrm{ILR}$ or $r_\mathrm{OLR}$, depending on the sign of $D^\prime_\mathrm{LR}$). Then we apply the Fush-Frobenius analysis in that region. Substituting the ansatz 
\begin{equation}
	\delta h = (x-r_{\rm LR})^\beta \sum_{k=0}^\infty a_k (r - r_\mathrm{LR})^k
\end{equation}
into equation (\ref{eq:master}), we find a system of equations for the coefficients $a_k$. The first two equations are
\begin{eqnarray}
	\beta(\beta - 2) a_0 &=& 0, \\
	(\beta^2 - 1) a_1 &=& \left[\frac{\tilde{m}A}{\tilde{\omega}} -
	\beta\partial_r\left(\frac{\sqrt{-g_3}\Sigma}{g_{rr}(u^t)^2}\right)\right]_\mathrm{LR} a_0.
\end{eqnarray}
The first equation implies $\beta = 0$ or 2, and the second one gives the first correction. Therefore, the solution up to the second order is
\begin{equation}
	\delta h = \delta h(r_\mathrm{LR})\left[1 - \left(\frac{\tilde{m}A}{\tilde{\omega}}\right)_\mathrm{L}(r - r_\mathrm{LR}) + 
	\mathcal{O}\left(|r - r_\mathrm{LR}|^2\right) \right].
\end{equation}
Therefore, the solution is regular at the Lindblad resonance. Using relation (\ref{eq:h2u}), we find that the radial-velocity perturbation is also 
finite at the Lindblad resonance and it is given by 
\begin{equation}
	\delta u_r(r_\mathrm{LR}) = \ii\left(\frac{\tilde{m}^2A^2 \delta h}{u^t D^\prime\tilde{\omega}}\right)_{\mathrm{LR}}.
\end{equation}

\subsection{Corotation resonance and relativistic vortensity}
\label{sec:corotation-resonance}

In the vicinity of the CR, $r=r_\mathrm{CR}$ where $\tilde{\omega}\approx 0$, the effective potential can be approximated as
\begin{equation}
	V_\mathrm{eff} \approx g_{rr}(u^t)^2 \frac{D}{\bar{c}_\mathrm{s}^2} - \frac{\tilde{m}A}{\tilde{\omega}}
	\partial_r\ln\left(\frac{\sqrt{-g_3}}{g_{rr}}\frac{\tilde{m}A\Sigma}{D u^t}\right).
\end{equation}
The wave equation then takes the form of the Whittaker equation
\begin{equation}
	\frac{\dd^2\psi}{\dd x^2} + \left[-\frac{1}{4} + \frac{\nu}{x + \ii\epsilon}\right]\psi = 0,
\end{equation}
with
\begin{equation}
	x = 2\int_{r_\mathrm{c}}^r \tilde{k} \dd r,
	\quad
	\psi = \tilde{k}^{1/2}\eta,
	\quad
	\tilde{k} = \sqrt{g_{rr}}u^t\frac{\kappa}{\bar{c}_\mathrm{s}}
\end{equation}
and
\begin{eqnarray}
	\nu &=& \frac{(1-\ell\Omega)rA\bar{c}_\mathrm{s}}{2q\kappa\Omega g_{rr} u^t}
	\partial_r\ln\left[\frac{g_{rr}}{\sqrt{-g_3}}\frac{\kappa^2 u^t}{(1-\ell\Omega)A\Sigma}\right],
	\\
	\epsilon &=& -\frac{2}{q}\frac{\omega_\mathrm{i}}{\omega_\mathrm{r}}\frac{r g_{rr} u^t \kappa}{\bar{c}_\mathrm{s}},
\end{eqnarray}
both evaluated at the corotation radius $r=r_{\rm CR}$. The $q$-parameter is defined as $q = -(\dd\ln\Omega)/(\dd\ln r)$. As in the non-relativistic case, the CR acts to amplify the p-modes, if it absorbs positive-energy waves. This happens when the Rossby wave zone lies on the right from the singularity, at $r>r_\mathrm{CR}$ , i.e.\ when $\nu > 0$.

The condition $\nu>0$ can be expressed in a more physical way using relativistic vorticity. For a perfect fluid the relativistic vorticity tensor is defined as
\begin{equation}
	\omega_{\mu\nu} = \nabla_\nu(w u_\mu) - \nabla_\mu(w u_\nu),
\end{equation}
where $w=(e+p)/\rho$ \citep[e.g.,][]{Teukolsky1998}. In thin Keplerian accretion discs $w\approx 1$ and the dominant component of $\omega_{\mu\nu}$ is
\begin{equation}
	\omega_{r\phi}=\frac{u_t\ell_{,r}}{1-\ell\Omega} = \frac{g_{rr}\kappa^2 u^t}{(1-\ell\Omega)A}.
\end{equation}
The relativistic vortensity can be introduced as
\begin{equation}
  \zeta = \frac{\omega_{r\phi}}{\sqrt{-g_3}\Sigma} = \frac{g_{rr}u^t\kappa^2}{\sqrt{-g_3}
  A\Sigma(1-\ell\Omega)}.
\label{eq:vortensity}\end{equation} 
In the Newtonian limit, this reduces to $\zeta\simeq \kappa^2/(2\Omega_\mathrm{K}\Sigma)$.  With this definition of the relativistic vortensity, the condition for the p-mode growth due to the CR takes the form
\begin{equation}
	\left(\frac{d}{dr}\ln\zeta\right)_{r=r_\mathrm{CR}} > 0.
\label{eq:criterion}
\end{equation}


\section{Disc p-modes in Kerr spacetime}
\label{sec:kerr}

\begin{figure*}
	\includegraphics[width=0.48\textwidth]{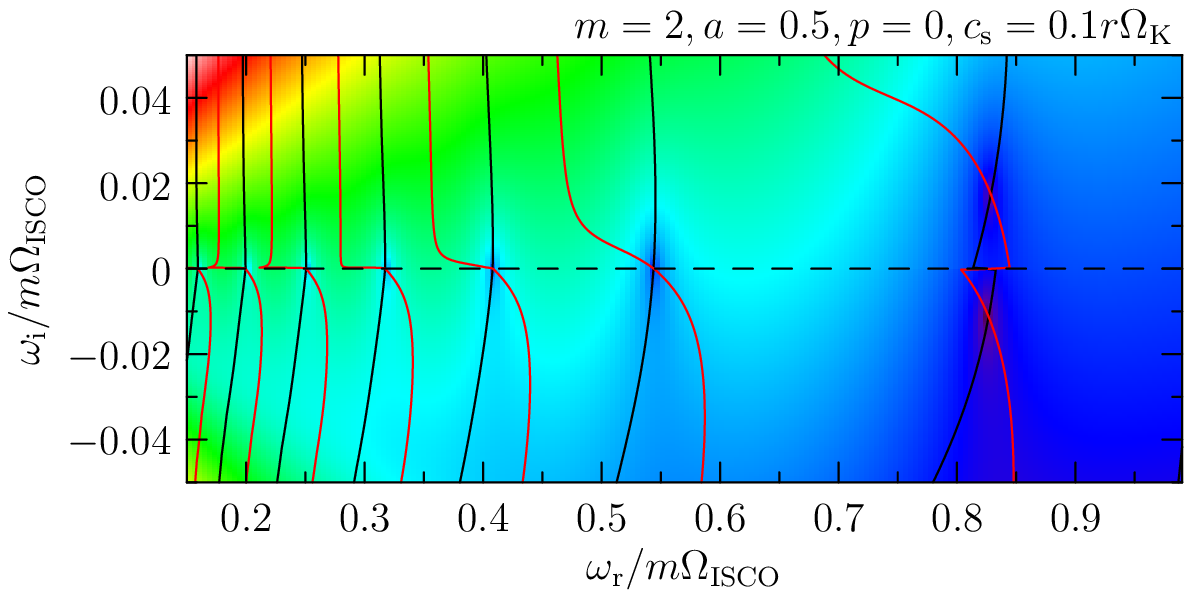}
	\hfill
	\includegraphics[width=0.48\textwidth]{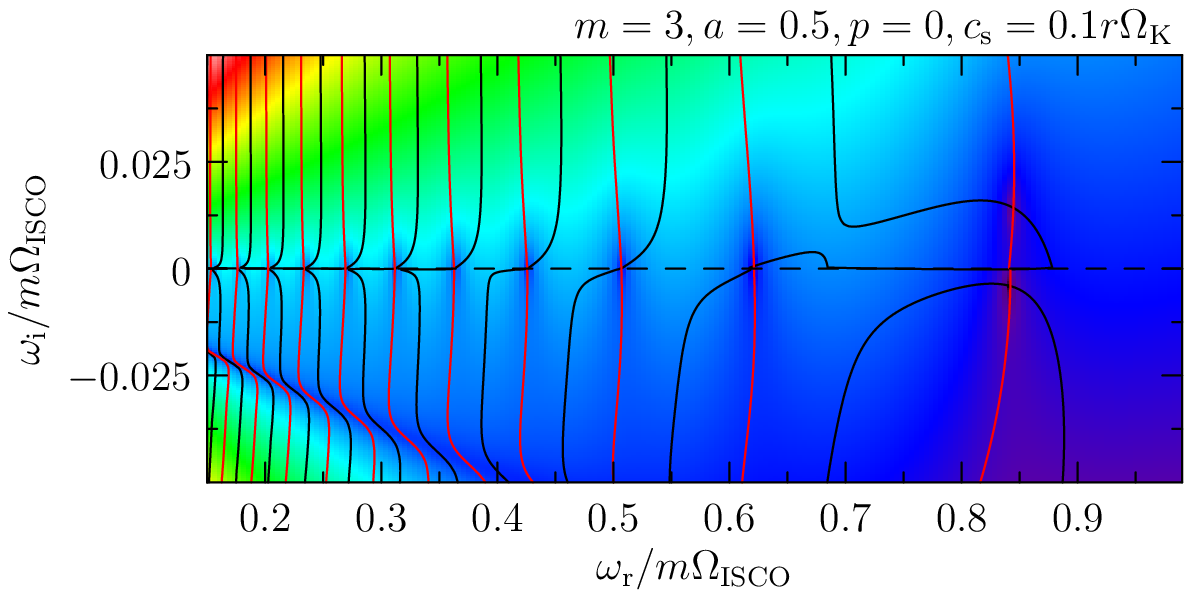}
	\\
	\includegraphics[width=0.48\textwidth]{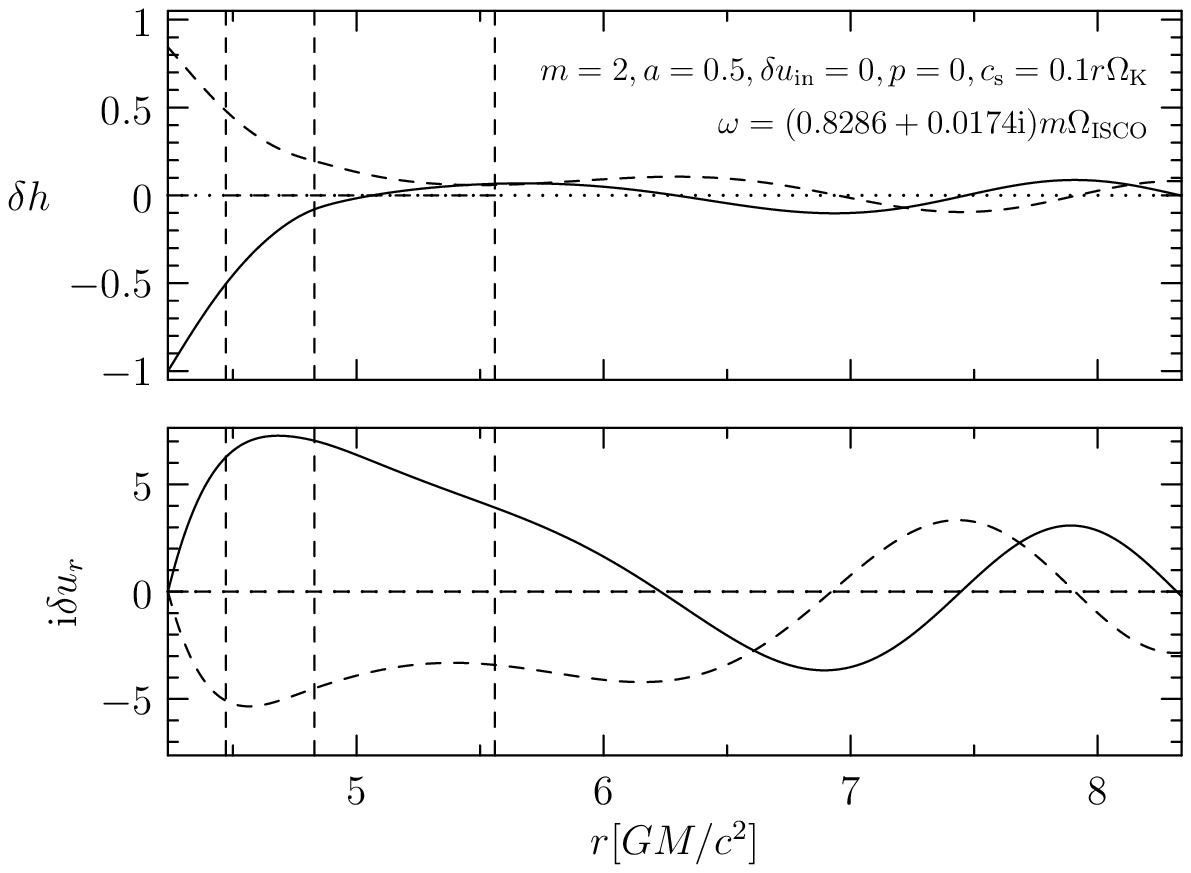}
	\hfill
	\includegraphics[width=0.48\textwidth]{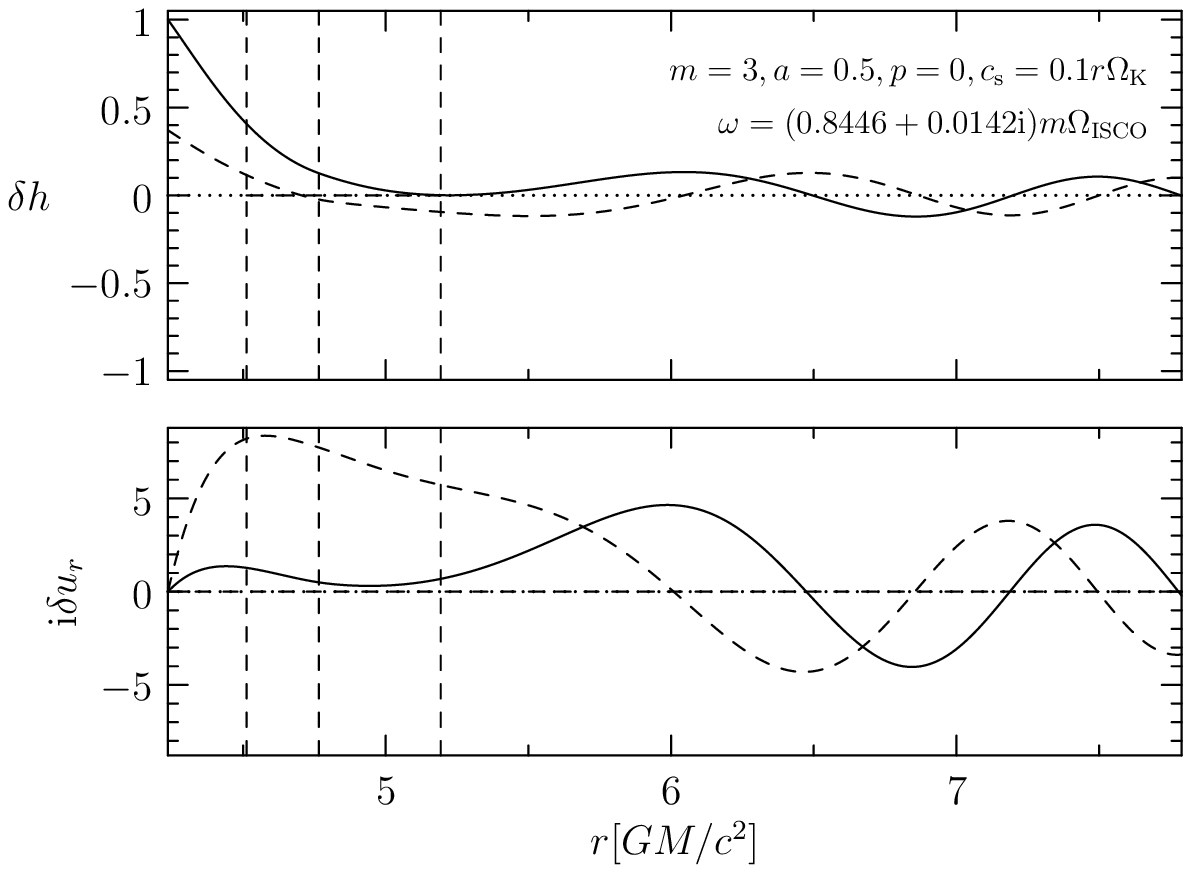}
	\caption{
		Top: the mismatch function $\Delta$ used in the shooting method for the $m=2$ (left) and $m=3$ (right) modes. The zeroes correspond to the eigenmodes of the disc. The red and black solid lines connect the points where the real or imaginary part of $\Delta$ vanishes. The colour corresponds to $|\Delta|^2$. Bottom: the eigenfunctions of the $m=2$ (left) and $m=3$ (right) modes. The solid and dashed lines correspond to the real and imaginary parts. The vertical dashed lines correspond to the inner Lindblad resonance, corotation resonance and the outer Lindblad resonance. The BH and disc parameters are $a=0.5$, $p=0$ and ${\bar c}_\mathrm{s} = 0.1r\Omega_\mathrm{K}$. The eigenmode frequencies are given by  $\omega/(m\Omega_{\rm ISCO})=0.8286+0.0174i$ (for $m=2$) and $0.8446+0.0142i$ (for $m=3$), where $\Omega_{\rm ISCO}$ is the Keplerian rotation frequency at the ISCO.
	}
	\label{fig:missmatch}
\end{figure*}


\subsection{Kerr metric}

The nonzero components of the Kerr metric in the equatorial plane are (in dimensionless units where $G=M=c=1$) given as
\begin{eqnarray}
	&&  g_{rr} = \frac{r^2}{\mathcal{R}^2},
	\quad
	g_{\theta\theta} = r^2,
	\quad
	g_{\phi\phi} = r^2 + a^2\left(1+\frac{2}{r}\right),
	\nonumber\\	
	&&  g_{tt} = -\left(1-\frac{2}{r}\right),
	\quad
	g_{t\phi} = -\frac{2a}{r},
\end{eqnarray}
where $\mathcal{R} = (r^2 - 2r + a^2)^{1/2}$. The Keplerian angular velocity and the specific angular momentum ($\ell_{\rm K}=-u_\phi/u_t$) are given by
\begin{equation}
	\Omega_\mathrm{K} = \frac{1}{a + r^{3/2}},
	\quad
	\ell_\mathrm{K} = \frac{a^2 - 2a r^{1/2} + r^2}{a - 2r^{1/2} + r^{3/2}}.
\end{equation}
The radial epicyclic frequency reads \citep[see, e.g.,][]{Okazaki+1987}
\begin{equation}
	\kappa = \left(1 - \frac{6}{r} + \frac{8a}{r^{3/2}} - \frac{3 a^2}{r^2}\right)^{1/2}\Omega_\mathrm{K}.
\end{equation}
The function $A=A_r$ follows from equation (\ref{eq:kappa-A}) or (\ref{eq:Ai}):
\begin{eqnarray}
	A =\frac{r^2-2r+a\sqrt{r}}{r^2-2r+a^2}\left(-\frac{2\Omega_\mathrm{K}}{r}\right).
\end{eqnarray}
Hence, in the limit of the nonrotating central object, we have $A = -2\Omega_\mathrm{K}/r$. Finally, the covariant and contravariant time components of the fluid four-velocity are
\begin{eqnarray}
	u_t &=& -\frac{r^{3/2} - 2 \sqrt{r} + a}{\sqrt{r^3 - 3 r^2 + 2 a r^{3/2}}}, \\
	u^t &=& \frac{a + r^{3/2}}{(r^3 - 3 r^2 + 2 a r^{3/2})^{1/2}}.
\end{eqnarray}


\subsection{Disk model}

For simplicity, we assume the following profiles of the disc surface
density and speed of sound\footnote{We denote the power-law density index by $p$, in order to be consistent with the notation of previous work of Lai \& Tsang (2009). However, it should not be confused with the symbol for the unperturbed pressure used in Section~\ref{sec:theory}.}:
\begin{equation}
  \Sigma \propto r^{-p}, 
  \quad
  \bar{c}_\mathrm{s} \propto r\Omega_\mathrm{K}.
\end{equation}

Fig.~\ref{fig:vortensity} shows some examples of the vortensity profile for these disc models. The vortensity is positive with a maximum at the radius $r=r_\mathrm{peak}$ that corresponds to the orbital frequency $\Omega_\mathrm{peak}=\Omega_\mathrm{K}(r_\mathrm{peak})$. According to the discussion in Section~\ref{sec:corotation-resonance}, only those p-modes with the corotation radius inside $r_\mathrm{peak}$, so that $(d\zeta/dr)_{r=r_{\rm CR}}$ is positive, can become overstable due to corotational wave absorption. This happens when the frequency of the mode exceeds $\omega_\mathrm{min}=m\Omega_\mathrm{peak}$. The right-hand panel of Fig.~\ref{fig:vortensity} shows the frequency $\omega_\mathrm{min}$ 
as a function of the BH spin for different values of the density index $p$. With increasing $p$, the position of the vortensity maxima moves
towards larger radii, decreasing the peak frequency $\Omega_\mathrm{peak}$. When $p\geq 3/2$ the vortensity is entirely increasing function of the radius and therefore $\omega_\mathrm{min}=0$.


\subsection{Boundary conditions}

We solve equations (\ref{eq:h}) and (\ref{eq:u}) with appropriate boundary conditions to determine the disc $p$-modes. At large radii ($r>r_{\rm OLR}$), we impose the outgoing-wave boundary condition,
\begin{equation}
	\left[\frac{d}{dr} - \ii k + \frac{1}{2}\frac{d}{dr}\left(\ln\frac{S}{k}\right)\right]\delta h = 0.
\end{equation}
In terms of the variables $\{\delta h, \delta u_r\}$, this reads
\begin{equation}
	\delta u_r + \frac{1}{D u^t}\left[k\tilde{\omega}-\ii\tilde{m} A - 
	\frac{\ii\tilde{\omega}}{2}\frac{d}{dr}\ln\frac{S}{k}\right]\delta h = 0.
\end{equation}

In our simplified disc models, the inner disc edge is located at the innermost stable circular orbit, $r_\mathrm{in} = r_\mathrm{ISCO}$. Following the previous works \citep{Lai+Tsang2009, Tsang+Lai2009c}, we adopt the condition of the vanishing radial-velocity perturbation at
$r=r_{\rm in}$:
\begin{equation}
	\delta u_r(r_\mathrm{in}) = 0.
	\label{eq:bc_ur}
\end{equation}
A significant uncertainty in calculating the disc p-modes is the inner boundary condition. The trapping of the p-modes between $r_{\rm in}$ and $r_{\rm ILR}$ requires that the inner disc boundary be partially reflective to incoming waves. In reality, the infall velocity of the disc gas increases rapidly as $r$ approaches $r_{\rm ISCO}$, leading to mode damping. The boundary condition (\ref{eq:bc_ur}) implicitly assumes perfect wave reflection at $r_{\rm in}$, and does not capture the effect of gas infall. Indeed, the absence of HFQPOs in the thermal state of BH X-ray binaries may be a consequence of mode damping due to gas infall. A more detailed study of this issue is needed (see Lai \& Tsang 2009 for an estimate). On the other hand, when a significant magnetic flux accumulates inside the ISCO, a magnetosphere may form around the BH \citep[e.g.,][]{Bisnovatyi-Kogan+Ruzmaikin1974, Bisnovatyi-Kogan+Ruzmaikin1976, Igumenshchev+2003, Rothstein+Lovelace2008, McKinney+2012}. Equation (\ref{eq:bc_ur}) may then serve as an approximate boundary condition at the magnetosphere-disc interface \citep[see][]{Tsang+Lai2009b, Fu+Lai2012}.


\subsection{Numerical procedure}

The eigenvalue problem is solved using the shooting method. For a given complex mode frequency $\omega$, we start from an outer disc radius, typically at $r_{\rm out} = 1.5 r_\mathrm{OLR}$, and integrate inwards equations (\ref{eq:u}) and (\ref{eq:h}) using the explicit embedded Runge--Kutta Prince--Dormand method. This way we obtain the values of the right-hand side solutions $\delta u_{r,\mathrm{R}}$ and $\delta h_\mathrm{R}$ at $r = r_\mathrm{in}$ and calculate the corresponding $(\eta_\mathrm{R}, \eta_\mathrm{R}^\prime)$. Similarly, we calculate the left-hand side solution $(\eta_\mathrm{L}, \eta_\mathrm{L}^\prime)$ from the inner boundary condition. The mismatch function is given by $\Delta(\omega) = \eta_\mathrm{L}\eta_\mathrm{R}^\prime - \eta_\mathrm{R}\eta_\mathrm{L}^\prime$. We than seek the roots of this function using the hybrid method. Examples of the mismatch functions are shown in the top panels of Fig.~\ref{fig:missmatch}. The bottom panels of Fig.~\ref{fig:missmatch} depict the eigenfunctions of the disc p-modes founded by this method; both the enthalpy perturbations $\delta h$ and the radial-velocity perturbations $\ii\delta u_r$ are shown.


\subsection{Frequencies of overstable p-modes: dependence on disc and black-hole parameters}

Using the method outlined in the previous subsections, we calculate the overstable p-modes for various disc parameters and BH spin.

\begin{figure*}
	\includegraphics[width=0.48\textwidth]{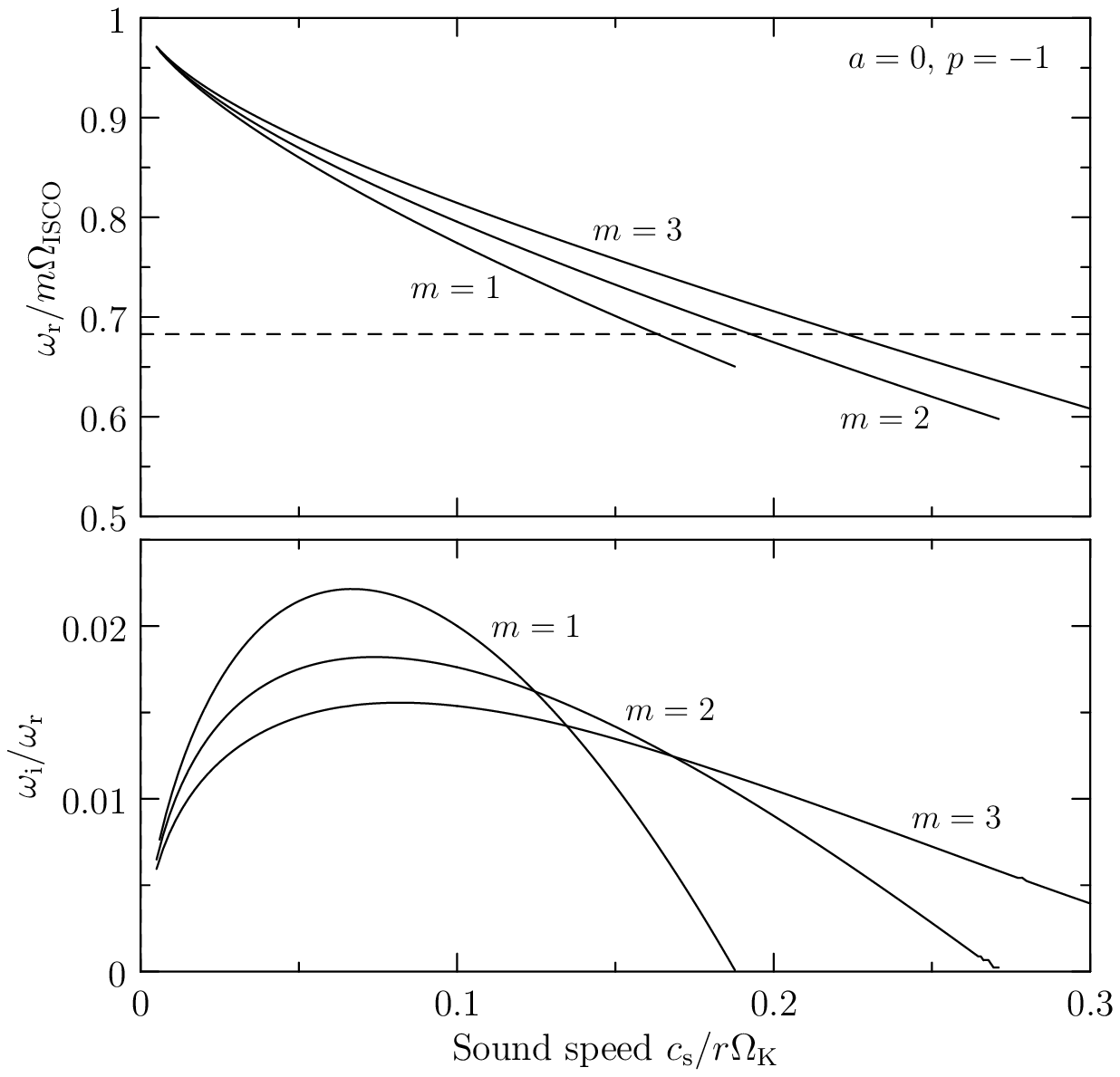}
	\hfill
	\includegraphics[width=0.48\textwidth]{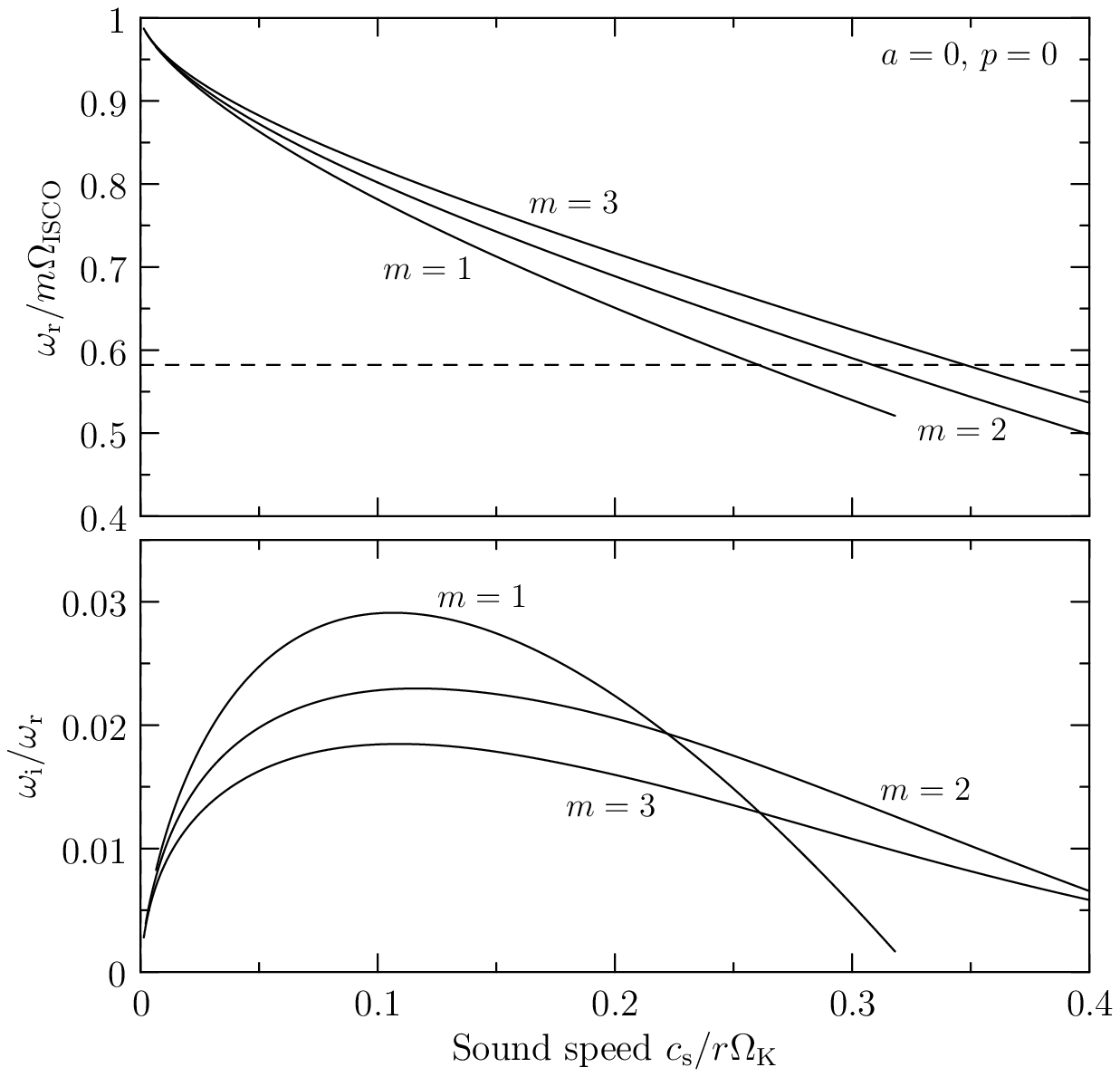}
	\caption{
		The eigenfrequencies of overstable p-modes as functions of the disc sound speed for increasing ($p=-1$, left-hand panels) and constant ($p=0$, right-hand panels) surface density profiles ($\Sigma\propto r^{-p}$). The horizontal dashed lines in the upper panels denote the minimal frequency for the corotational wave absorption to amplify the oscillations.
	}
	\label{fig:trace-c}
\end{figure*}

(i) {\it Dependence on the sound speed:} Fig.~\ref{fig:trace-c} shows the complex eigenfrequencies $\omega=\omega_\mathrm{r} + \ii \omega_\mathrm{i}$ of p-modes as functions of the disc sound speed. The left-hand panels correspond to $p=-1$ (increasing surface density with radius) and the right-hand panels correspond to $p=0$ (constant surface density), both for the BH spin parameter $a=0$. In the top panels, the horizontal dashed lines denote the orbital frequency at the vortensity maximum, corresponding to the minimal frequency $\omega_{\rm min}$ of the p-modes for which wave absorption at the CR leads to mode growth (see Fig.~\ref{fig:vortensity}). Note, however, that even when the mode frequency is slightly below $\omega_{\rm min}$, the p-modes can still be overstable; this is because the ``transmitted'' outgoing waves (in the region $r>r_{\rm OLR}$) always lead to over-reflection and mode growth \citep[see][]{Tsang+Lai2008}.

\begin{figure}
	\includegraphics[width=0.49\textwidth]{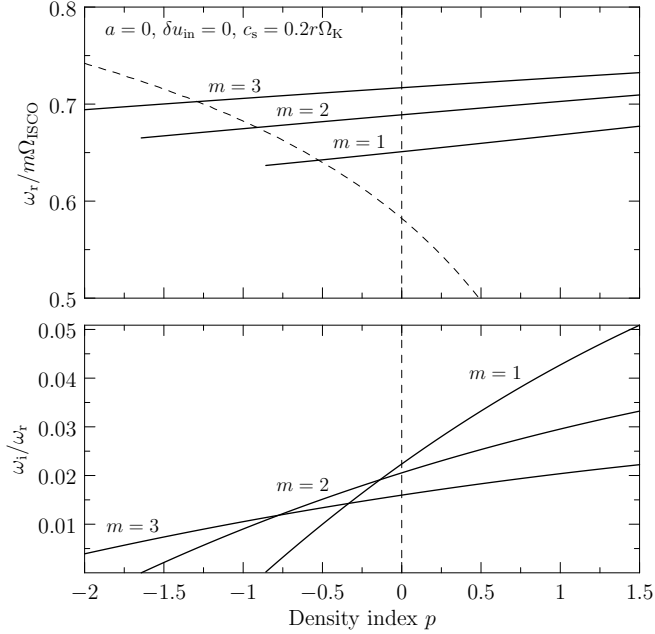}
	\caption{
		The eigenfrequencies of the unstable p-modes as functions of the density index $p$. The bent dashed line on the top panel corresponds to the minimal frequency for the corotational wave absorption to amplify the oscillations.
	}
	\label{fig:trace-p}
\end{figure}

(ii) {\it Dependence on the surface density profile:} Fig.~\ref{fig:trace-p} depicts the dependence of the p-mode eigenfrequency on the surface density distribution index $p$. The behaviour of the frequencies with changing $p$ is qualitatively similar to the case of the pseudo-Newtonian discs \citep[cf. fig.~7 (left) of][]{Lai+Tsang2009}.

\begin{figure}
  \includegraphics[width=0.49\textwidth]{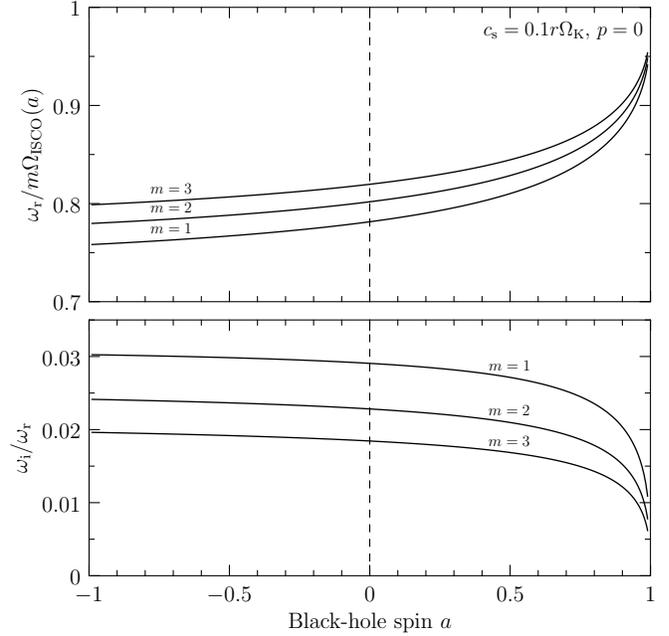}
  \caption{The eigenfrequencies of the unstable p-modes as functions the black hole spin parameter $a$.}
  \label{fig:trace-a}
\end{figure}

(iii) {\it Dependence on the BH spin:} Fig.~\ref{fig:trace-a} depicts the dependence of the p-mode eigenfrequency on the BH spin parameter $a$. Most strikingly, we see that as $a$ approaches unity, the real mode frequency increases almost to $m\Omega_\mathrm{ISCO}$ (where $\Omega_\mathrm{ISCO}$ is the orbital frequency at the ISCO). 

\begin{figure}
	\includegraphics[width=0.49\textwidth]{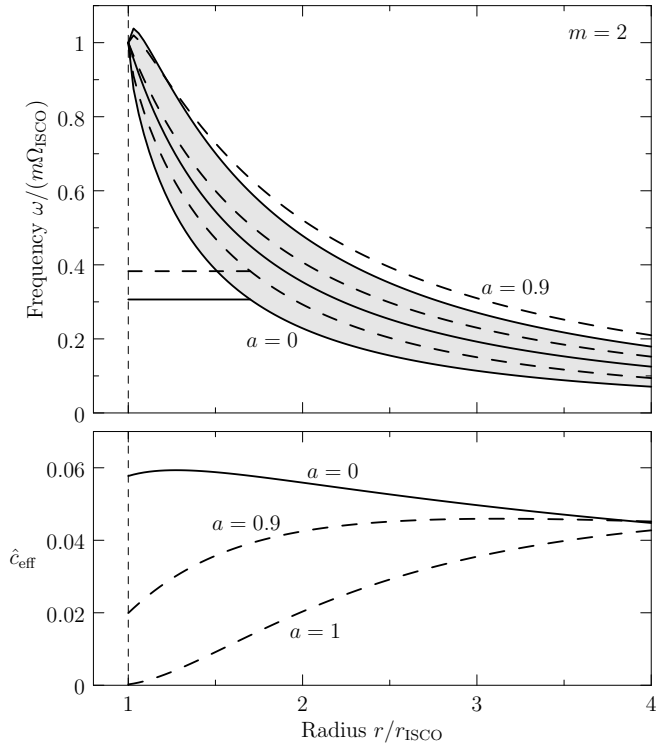}
	\caption{
		Upper panel: propagation diagram for the $m=2$ p-mode oscillations. The solid and dashed curves correspond to the black hole spin $a=0$ and 0.9, respectively. From left to right: the three curves are $\Omega-\kappa/2$, $\Omega$ and $\Omega+\kappa/2$, all in units of $\Omega_{\rm ISCO}$.  The two horizontal lines give the mode frequency $\omega_r$ in units of $m\Omega_{\rm ISCO}$. The p-modes are trapped between $r_{\rm in}=r_{\rm ISCO}$ and $r_{\rm ILR}$. Lower panel: the effective sound speed  $\hat{c}_\mathrm{eff}(r/r_\mathrm{ISCO})$ as a function of $r$ for three different BH spin parameters.
	}
	\label{fig:propagation}
\end{figure}

\begin{figure}
	\includegraphics[width=0.49\textwidth]{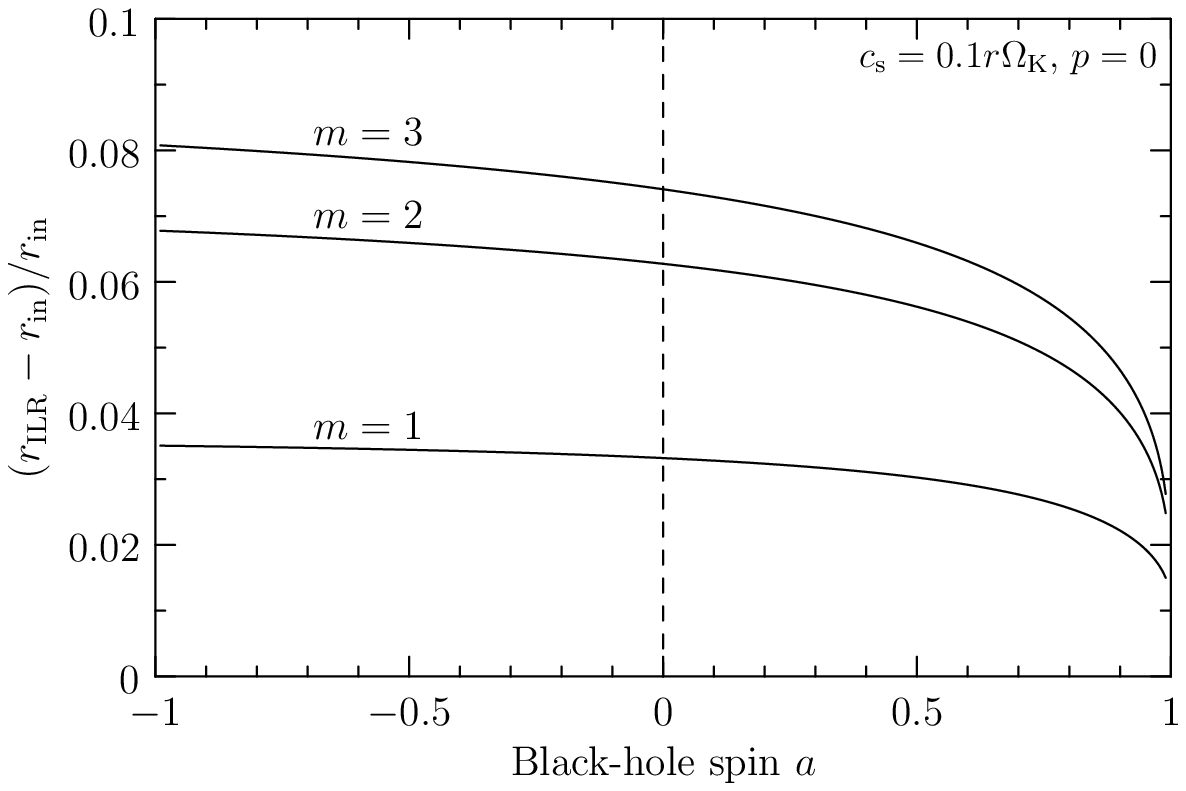}
	\hfill
	\includegraphics[width=0.49\textwidth]{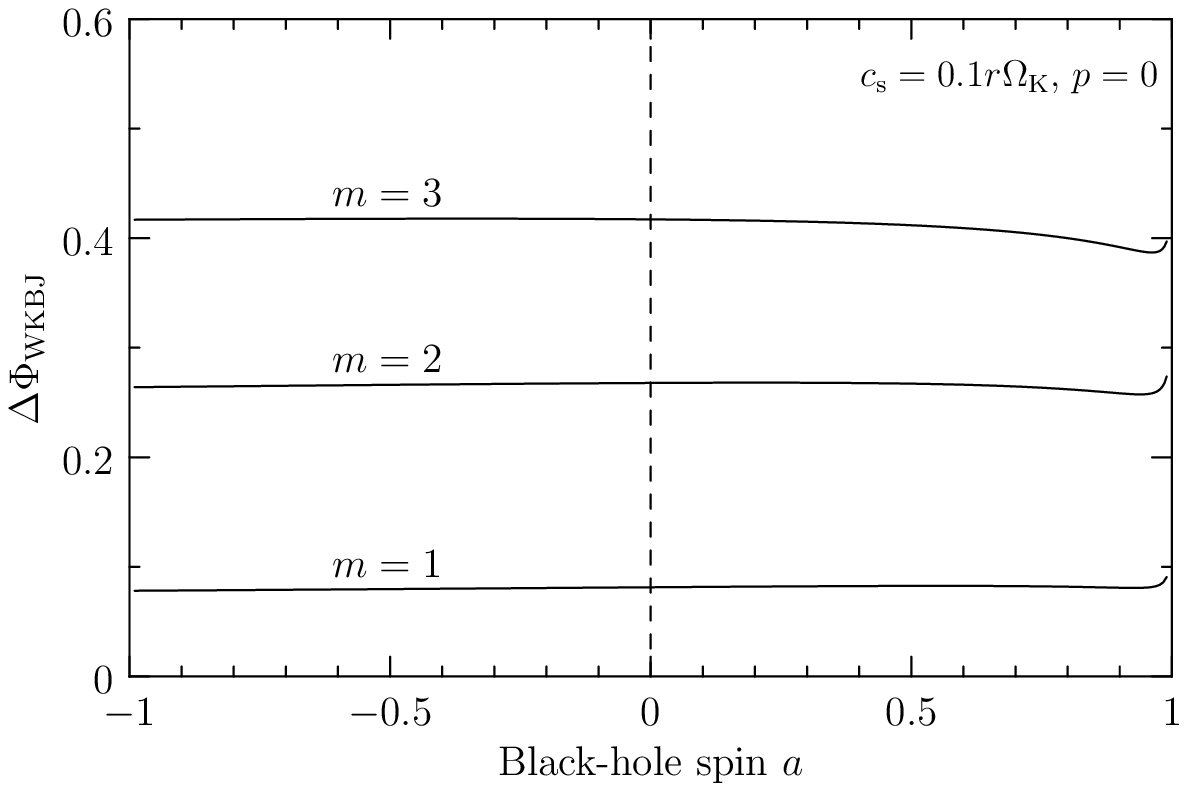}
	\caption{
		The width of the trapping region (upper panel) and the WKBJ phase difference (lower panel) for the most unstable p-modes as a function of the BH spin parameter.
	}
	\label{fig:trapping-phase}
\end{figure}

The dependence of the mode frequency $\omega_r$ on $a$ can be understood from the wave propagation diagrams shown in Fig.~\ref{fig:propagation} and the WKBJ analysis. Since the p-modes are trapped between the disc inner edge $r_{\rm in}$ and the inner Lindblad resonance $r_{\rm ILR}$, 
the mode frequency is determined approximately by the Sommerfeld quantization condition:
\begin{eqnarray}
	\Delta\Phi_\mathrm{WKBJ} &=& \int_{r_\mathrm{in}}^{r_\mathrm{ILR}}
	k\, \dd r = \int_1^{r_\mathrm{ILR}/r_\mathrm{in}}
	\frac{1}{\hat{c}_\mathrm{eff}}\left(\frac{\sqrt{-D}}{\Omega_\mathrm{in}}\right)\dd x
	\nonumber \\
	&=& n\pi + \varphi,
\end{eqnarray}
where the WKBJ radial wavenumber $k$ is given by equation (\ref{eq:k}), and $x=r/r_\mathrm{in}$, $n$ is an integer, $\varphi$ is of the order of unity and 
\begin{equation}
	\hat{c}_\mathrm{eff} = \frac{1}{g_{rr}^{1/2}u^t}\frac{\bar{c}_\mathrm{s}}{r_\mathrm{in}\Omega_\mathrm{in}}
\end{equation}
is the effective sound speed. Clearly, when ${\bar c}_\mathrm{s}$ decreases, the eigenfrequency $\omega_r$ must increase in order to decrease $r_{\rm ILR}/r_{\rm ISCO}$, so that the same amount of $\Delta\Phi_{\rm WKBJ}$ can `fit' in the mode trapping zone. This explains the behaviour of $\omega_r$ as a function of ${\bar c}_\mathrm{s}$ in Fig.~\ref{fig:trace-c}. On the other hand, as the BH spin $a$ increases, the inner disc radius $r_{\rm ISCO}/M$ decreases, leading to larger $g_{rr}^{1/2} u^t$ and smaller effective sound speed ${\hat c}_{\rm eff}$ (see the lower panel of Fig.~\ref{fig:propagation}). Indeed, $g_{rr}^{1/2} u^t$ diverges to infinity and ${\hat c}_{\rm eff}$ approaches zero as $r\rightarrow r_\mathrm{ISCO}$ and $a\rightarrow 1$. Therefore, with increasing $a$, the mode frequency $\omega_r$ (in units of $\Omega_{\rm ISCO}$) must increase (lowering the $\sqrt{-D}$ factor and reducing the size of the trapping region) in order to keep $\Delta\Phi_\mathrm{WKBJ}$ roughly constant and satisfy the quantization condition (see Fig.~\ref{fig:trapping-phase}). This explains the behaviour of $\omega_r$ as a function of $a$ seen in the upper panel of Fig.~\ref{fig:trace-a}.

\begin{figure}
	\includegraphics[width=0.49\textwidth]{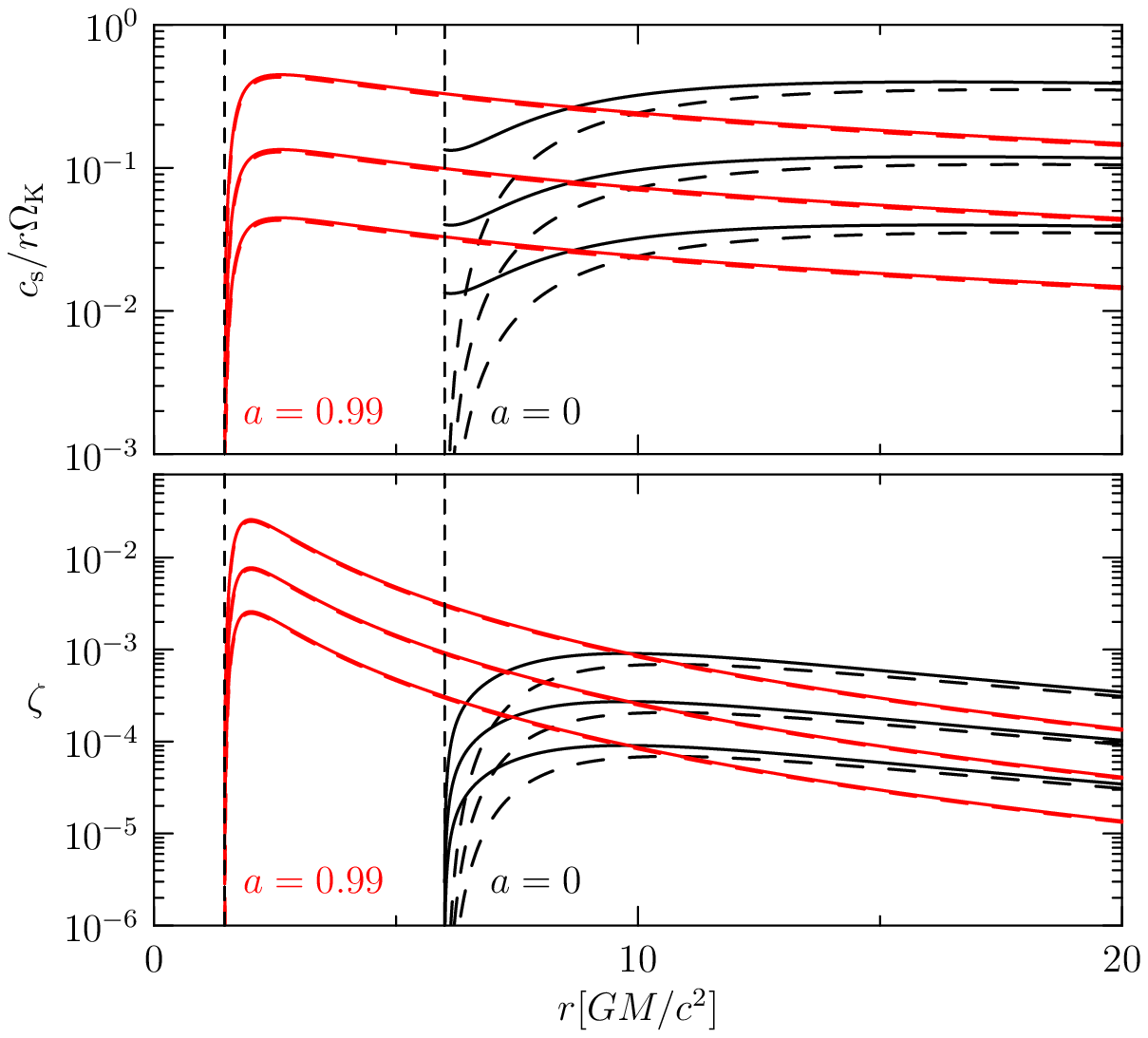}
	\caption{
		The speed of sound (top) and vortensity (bottom) profiles for the radiation pressure dominated Novikov-Thorne disc models. Two sets of curves are shown: the left-most red curves that terminate at $r=1.46 M$ correspond to the black hole spin $a=0.99$ while the black curves terminating at $r=6M$ are for the Schwarzschild BH with $a = 0$. These radii correspond to the ISCOs and are shown by the vertical dashed lines. Different curves in each set show behaviour for different Eddington ratios: $L/L_\mathrm{Edd} = 0.1$, 0.3, and 1 (from bottom to top). In addition, the solid and dashed lines correspond to zero ($F_\ast=1$) and a small but nonzero ($F_\ast = 0.95$) torque at the inner disc boundary.
	}
	\label{fig:nt-profiles}
\end{figure}


\subsection{Overstable p-modes in Novikov--Thorne discs}
\label{sec:nt}

\begin{figure*}
	\includegraphics[width=0.32\textwidth]{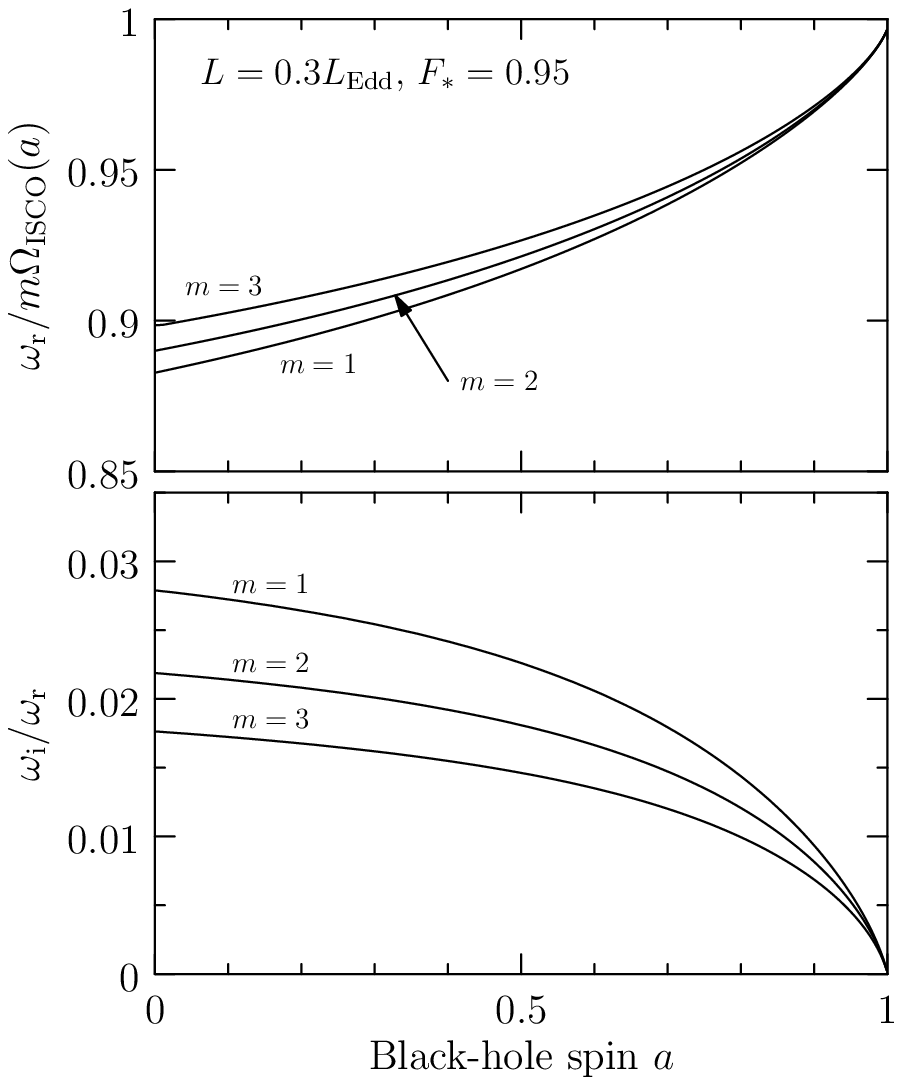}
	\hfill
	\includegraphics[width=0.32\textwidth]{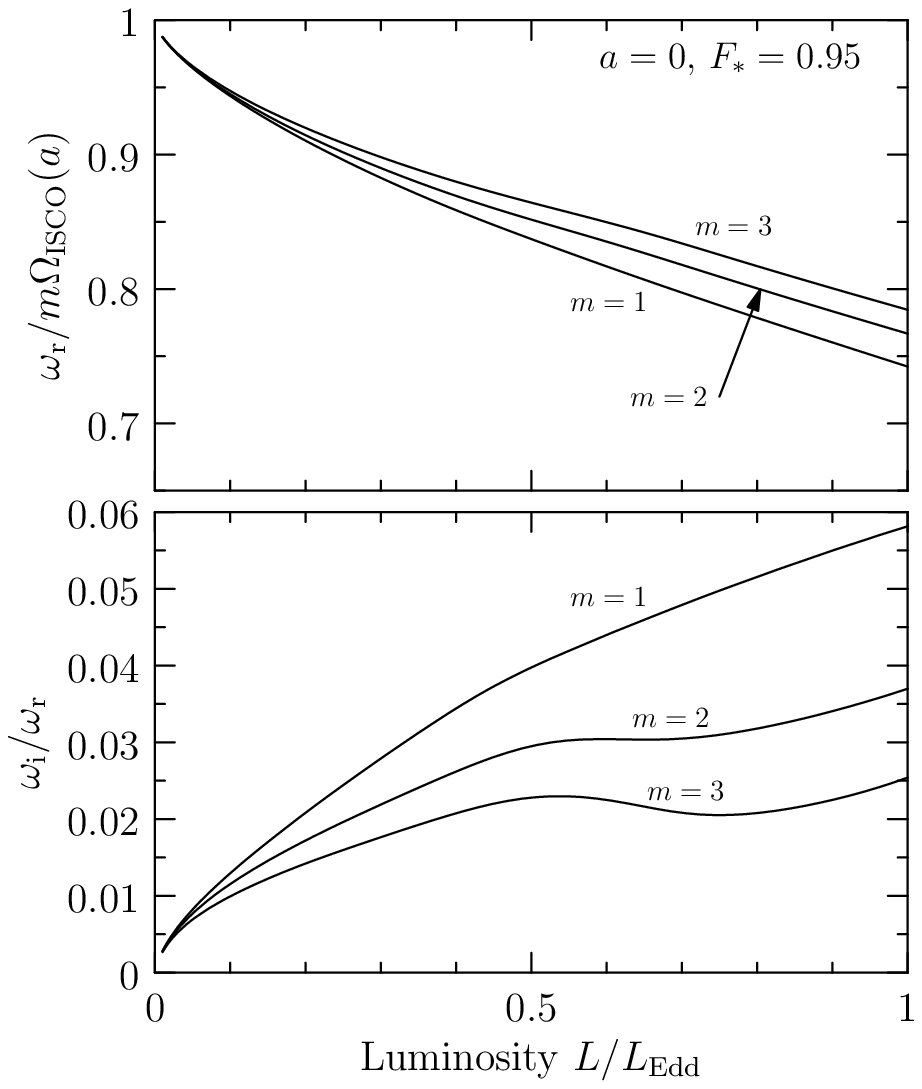}
	\hfill
	\includegraphics[width=0.32\textwidth]{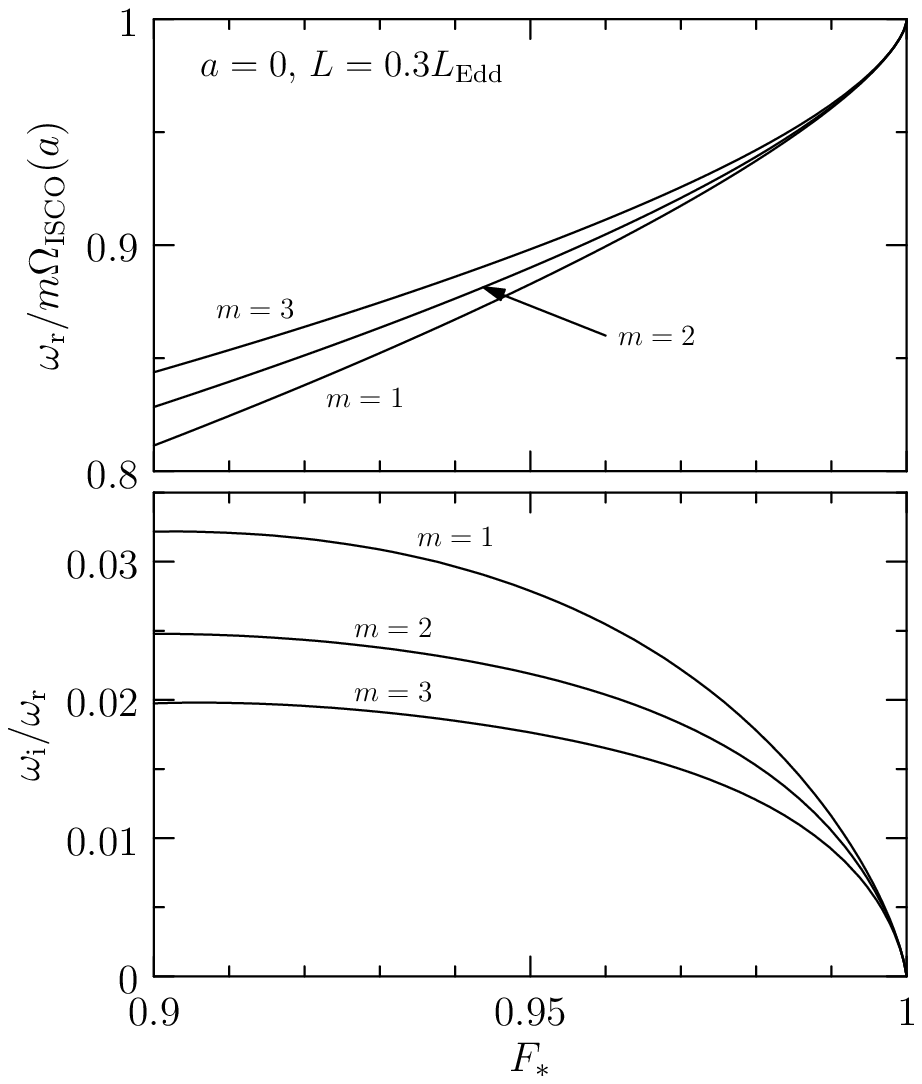}
	\caption{
		The eigenfrequencies of the overstable p-modes in radiation pressure dominated Novikov-Thorne discs as functions of black hole spin parameter $a$ (left), luminosity $L$ (middle; in units of the Eddington luminosity) and the torque parameter $F_\star$  at the inner boundary (right). The zero-torque inner boundary condition corresponds to $F_\ast = 1$.
	}
	\label{fig:nt}
\end{figure*}

Having studied the basic properties of corotation instability in our parametrized disc models, we now turn to Novikov--Thorne $\alpha$ disc models. We assume that the p-modes are trapped in the inner regions of the disc where the flow is dominated by radiation pressure. The density and sound speed profiles are then given as \citep{Novikov+Thorne1973, Ortega-Rodriguez+2008}
\begin{eqnarray}
	\Sigma &\propto& \left(L/L_\mathrm{Edd}\right)^{-1}(r/r_\mathrm{g})^{3/2}\eta(a)
	\nonumber\\	&\phantom{=}&
	\times\mathcal{A}^{-2}\mathcal{B}^3\mathcal{C}^{1/2}
	\mathcal{E}\left[\mathcal{Q} + \delta\mathcal{Q}\right],
	\\
	\frac{c_\mathrm{s}}{r\Omega} &=& 1.3 \left(L/L_\mathrm{Edd}\right)
	\left(r/r_\mathrm{g}\right)^{-1} \eta(a)^{-1}
	\nonumber\\ &\phantom{=}&
	\times\mathcal{A}\mathcal{B}^{-1}\mathcal{D}^{-1/2}\mathcal{E}^{-1/2}
	\left[\mathcal{Q} + \delta\mathcal{Q}\right],
\end{eqnarray}
where $L$ and $L_\mathrm{Edd}$ are the total luminosity of the disc and Eddington luminosity, $\eta(a)$ is the efficiency that relates the luminosity and the accretion rate as $\dot{M}=L/(\eta c^2)$, $r_\mathrm{g}=GM/c^2$ is the gravitational radius, and $\mathcal{A}$, $\mathcal{B}$, $\mathcal{C}$, $\mathcal{D}$, $\mathcal{E}$, $\mathcal{F}$ and $\mathcal{Q}$ are functions of $r$ and $a$ as defined in \citet{Novikov+Thorne1973} and take values of unity far from the BH. Following \citet{Ortega-Rodriguez+2008}, we introduce a function $\delta\mathcal{Q}$ to allow for a non-zero torque at the inner boundary of the disc, at $r = r_\mathrm{in} = r_\mathrm{ISCO}$:
\begin{eqnarray}
	\delta Q = (1 - F_\ast)\mathcal{B}_\mathrm{in}\mathcal{C}_\mathrm{in}^{1/2}\mathcal{B}^{-1}\mathcal{D}^{1/2}\mathcal{F},
\end{eqnarray} 
where the index `in' denotes evaluation at $r_\mathrm{in}$ and $F_\ast$ is the fraction of the inflowing specific angular momentum at $r_\mathrm{in}$ that is absorbed by the BH. There is a torque at $r_\mathrm{in}$ unless $F_\ast$ = 1.

The sound speed and vortensity profiles are shown in Fig.~\ref{fig:nt-profiles} for Schwarzschild ($a=0$) and nearly maximally rotating ($a=0.99$) BH. The sound speed is vanishing at the inner disc edge when there is no torque at the disc inner boundary ($F_\ast =1$), while it remains finite for $F_\ast<1$.

The three panels of Fig.~\ref{fig:nt} show the behaviour of the unstable $m=1$, $m=2$ and $m=3$ mode frequencies with changing BH spin, luminosity and torque at the inner boundary. The spin dependence is very similar to the one obtained in the previous subsection for the parametrized disc models. Indeed, as discussed there, it is mainly the geometrical factor $g_{rr}^{1/2} u^t$ that leads to the increase of $\omega_\mathrm{r}/(m\Omega_\mathrm{in})$ as $a\rightarrow 1$, quite independently on the adopted disc models.

Similarly, the behaviour of the modal frequencies with changing luminosity agrees qualitatively with the results of the previous subsection. With increasing $L$, the magnitude of the sound speed grows and the real parts of the eigenfrequencies must decrease in order to increase the trapping regions of the modes. Higher sound speed also facilitates the tunneling of the waves to the Rosby wave zone, increasing the growth-rates of the modes. The subsequent decrease of $\omega_i/\omega_r$ observed in Fig.~\ref{fig:trace-c} may in principle occur at still higher luminosities, although they are clearly out of the range of applicability of the Novikov--Thorne models.

Finally, the behaviour of the mode frequencies as a function of the inner torque parameter $F_\ast$ is similar to the spin dependences in the leftmost panel of Fig~\ref{fig:nt}. The fact that $\omega_r$ increases to $m\Omega_\mathrm{in}$ as $F_\ast\rightarrow 1$ is connected to vanishing sound speed at the inner edge of the disc.  


\section{Discussion}
\label{sec:discussion}

As noted in Section 1, the origin of HFQPOs in BH X-ray binaries is currently unknown. Significant progress in numerical magnetohydrodynamic (MHD) (including GRMHD) simulations of BH accretion flows has been made in the past decade, but much work remains to capture the complex phenomenology of BH X-ray binaries. Several recent simulations have revealed quasi-periodic variabilities of various fluid variables, but the connection of these variabilities to the observed HFQPOs is far from clear \citep[e.g.,][]{Henisey+2009, ONeill+2011, Dolence+2012, McKinney+2012, Shcherbakov+McKinney2013}. It is likely that future progress in our understanding of HFQPOs would require a combination of high-quality data (as may be provided by the large observatory for X-ray timing, LOFT; see \citet{Feroci+2012}, full numerical simulations and semi-analytic studies to extract the underlying physics.

In this paper, we have studied the effect of CR on spiral wave modes in general relativistic discs around BHs. In Newtonian theory, it is known that wave energy can be absorbed at the CR; depending on the sign of the disc vortensity gradient, $(\dd/\dd r)(\kappa^2/2\Omega\Sigma)$, such corotational wave absorption can lead to the growth of spiral waves \citep[for other applications where CR and vortensity gradient play an important role;]{Tsang+Lai2008, Tsang+Lai2009c, Goldreich+Tremaine1979, Narayan+1987, Papaloizou+Pringle1987, Lovelace+1999, Meheut+2012}.  Our formulation developed in Section 2 generalizes the theory of CR to fully relativistic discs [see equations~(\ref{eq:vortensity}) and (\ref{eq:criterion})]. We calculated the overstable p-modes trapped in the inner-most region of a BH accretion disc based on simple disc models parametrized by disc sound speed and surface density profile. This approach helped us to isolate and understand various effects that determinate frequencies and growth rates of the modes. Later, we also studied corotation instability for Novikov \& Thorne disc models.

\begin{figure}
	\includegraphics[width=0.49\textwidth]{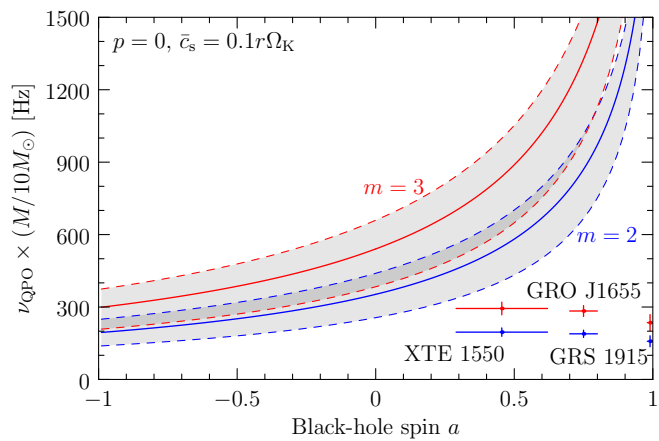}
	\caption{
		The frequency of overstable p-modes rescaled to the BH mass of $10\,M_\odot$ as a function of the BH spin parameter. The two solid curves (blue and red) correspond to the computed $m=2$ and 3 mode frequencies for discs with $p=0$ and $c_\mathrm{s}=0.1 r\Omega_\mathrm{K}$. The surrounding dashed areas are delimited by the frequencies $m\Omega_\mathrm{ISCO}$ and $m\Omega_\mathrm{peak}$    for both $m=2$ and $m=3$ cases. Hence, they correspond to the regions where the corotational wave absorption acts to amplify the waves.
	}
	\label{fig:qpos}
\end{figure}

Our selected numerical results of the frequency and growth rate of p-modes for different BH and disc parameters are presented in Figs.~\ref{fig:trace-c} -\ref{fig:trace-a} and \ref{fig:nt}. In units of $\Omega_{\rm in} =\Omega_{\rm ISCO}$ (the rotation frequency at the inner disc radius $r_{\rm in}=r_{\rm ISCO}$), the dimensionless mode frequency $\hat\omega_r=\omega_r/(m\Omega_{\rm in})$ ranges from $0.6$ to $1$
for the lowest-order (highest-frequency) p-modes, and its precise value depends on $a$ (BH spin), $p$ (the surface density profile index as in $\Sigma\propto r^{-p}$), ${\bar c}_\mathrm{s}$ (sound speed) and $m$ (the azimuthal mode number). In physical units, the frequency of the p-mode with azimuthal number $m$ can be written as
\begin{equation}
	\nu_m\simeq 220\,m F(a)
	\,{\hat\omega_r}\,
	\left(\!{10M_\odot\over M}\!\right)\,{\rm Hz},
\end{equation}
with 
\begin{equation}
	F(a)\equiv {\Omega_{\rm in}(a)\over \Omega_{\rm in}(a=0)}.
\end{equation}
Fig.~\ref{fig:qpos} gives an example of $\nu_m$ as a function of the BH spin $a$ for a particular disc model. 

Compared to the observations of HFQPOs in BH X-ray binaries \citep{Remillard+McClintock2006, Belloni+2012}, our computed frequencies of $m=1$ p-modes are generally consistent with the observed values, given the measurement of the BH mass and the constraint on the BH spin. However,
if we interpret the observed harmonic pairs as the $m=2$ and 3 p-modes, then our computed frequencies are too high. For example, XTE J1550-564 (mass $M=9.1\pm 0.61 M_\odot$) has two HFQPOs at 184 and 276~Hz; this would require ${\hat\omega}_r =\omega_r/(m\Omega_{\rm in})\simeq 0.38/F(a)$, or $\hat\omega_r\le 0.38$ if $a>0$. Similarly, for GRO J1655-40 (mass $M=6.3\pm 0.3M_\odot$), to explain the observed HFQPOs at 300 and 450~Hz with the $m=2$ and 3 p-modes would require ${\hat\omega}_r\simeq 0.43/F(a)$. Note that the BH spin parameters for the above two systems have been constrained using the continuum spectrum fitting method to be $a=0.34\pm 0.24$ and $0.7\pm 0.1$, respectively \citep[see][and references therein]{Narayan+McClintock2012}. The discrepancy becomes most severe for GRS 1915+105 ($M=14\pm 4.4M_\odot$), which may have two pairs of HFQPOs (41 and 67~Hz, 113 and 168~Hz), and whose spin parameter has been constrained to be $a>0.975$ using the continuum fitting method.

Thus, it appears that for the simplest disc models considered in this paper, non-axisymmetric p-modes have frequencies that are too high compared to the observation of HFQPOs in BH X-ray binaries. A number of effects or complications may decrease the theoretical p-mode frequencies. For example, a higher disc sound speed leads to lower mode frequencies, and a steeper surface density profile (larger $p$ in $\Sigma\propto r^{-p}$) reduces $\omega_{\rm min}$, the minimum frequency for the corotational wave absorption to amplify the mode.  Magnetic fields may also play an important role. Although toroidal disc magnetic fields tend to suppress the corotational instability \citep{Fu+Lai2011}, large-scale poloidal fields threading the disc can enhance the instability and reduce the p-mode frequency \citep[see][]{Tagger+Pellat1999, Tagger+Varniere2006, Yu+Lai2013}. Since episodic jets are produced in the same spectral state (the `intermediate state') where HFQPOs are observed in BH X-ray binaries, such large-scale poloidal magnetic fields may indeed be present. Finally, our calculations presented in this paper assume that the inner disc radius coincides with the ISCO. This may not be the case during the `intermediate state' \citep[e.g.,][]{Done+2007, Oda+2010}. As noted before, a major uncertainty in calculating the disc p-modes is the inner disc boundary condition. When magnetic fields advect inwards in the accretion disc and accumulate around the BH \citep[e.g.][]{Bisnovatyi-Kogan+Ruzmaikin1974, Bisnovatyi-Kogan+Ruzmaikin1976, Igumenshchev+2003, Rothstein+Lovelace2008}, the inner disc radius $r_{\rm in}$ may be larger than $r_{\rm ISCO}$. This reduces $\Omega_{\rm in}$ relative to $\Omega_{\rm ISCO}$, leading to lower $p$-mode frequencies. Significant magnetic fields also make the disc sub-Keplerian \citep[as in the models of][]{McKinney+2012}, therefore changing the mode frequencies. These issues should be addressed in future studies.

In this paper, we have not addressed how disc oscillations may manifest as X-ray flux variability. We do note that, for the same amplitudes, low-$m$ modes are expected to be more `visible' than high-$m$ modes. Also, the large-scale magnetic field mentioned in the last paragraph may also play an important role: it can `channel' the disc oscillation to the corona and therefore produce variability in the hard X-ray flux (as observed in HFQPOs).


\section*{Acknowledgements}
This work has been supported in part by NSF grants AST-1008245 and AST-1211061, NASA grant NNX12AF85G, and Czech grants M100031202, ME~09036 and P209/11/2004.


\bibliographystyle{mn2e}
\bibliography{corot}

\label{lastpage}
\end{document}